\documentclass[11pt]{article}

\usepackage{tikz}

\usepackage{amsmath,amsfonts,amsthm,amssymb,color}

\usepackage{fullpage}

\newtheorem{theorem}{Theorem}[section]
\newtheorem{corollary}[theorem]{Corollary}
\newtheorem{lemma}[theorem]{Lemma}
\newtheorem{definition}[theorem]{Definition}

\newtheorem{remark}[theorem]{Remark}

\newcommand{\eps}{\varepsilon}

\usepackage{amsmath,amssymb,amsthm,amsfonts,latexsym,bbm,xspace,graphicx,float,mathtools,mathdots}
\usepackage{braket,caption,subcaption,ellipsis,xcolor,textcomp,hhline,pifont,combelow}
\definecolor{newblue}{rgb}{0.19, 0.55, 0.91}
\usepackage[colorlinks,citecolor=newblue,bookmarks=true,pagebackref=true, urlcolor=blue, linkcolor=blue, linktoc=page]{hyperref}
\usepackage[nameinlink]{cleveref}
\usepackage[letterpaper,margin=1in]{geometry}
\usepackage{inconsolata}
\usepackage{diagbox}
\usepackage{multicol}
\usepackage{hhline}

\usepackage{url}
\usepackage{amsmath}
\usepackage{mathtools}
\usepackage{amssymb}
\usepackage{graphicx}
\usepackage{stmaryrd}
\usepackage{arydshln}
\usepackage{multirow}
\usepackage[ruled,linesnumbered]{algorithm2e}

\newcommand{\Alg}{\ensuremath\mathsf{Alg}}
\newcommand{\cE}{\ensuremath\mathcal{E}}
\newcommand{\cF}{\ensuremath\mathcal{F}}
\newcommand{\cQ}{\ensuremath\mathcal{Q}}

\usepackage{amsmath,amsfonts,bm}

\def\1{\bm{1}}

\DeclareMathAlphabet{\mathsfit}{\encodingdefault}{\sfdefault}{m}{sl}
\SetMathAlphabet{\mathsfit}{bold}{\encodingdefault}{\sfdefault}{bx}{n}

\def\gA{{\mathcal{A}}}

\def\gD{{\mathcal{D}}}

\def\gN{{\mathcal{N}}}

\DeclareMathOperator*{\E}{\mathbb{E}}

\newcommand{\R}{\mathbb{R}}

\renewcommand{\tilde}{\widetilde}
\renewcommand{\bar}{\overline}
\newcommand{\eqdef}{:=}

\newcommand{\cD}{\mathcal{D}}
\newcommand{\cA}{\mathcal{A}}
\newcommand{\cS}{\mathcal{S}}
\newcommand{\bX}{\mathbf{X}}
\newcommand{\bY}{\mathbf{Y}}

\newcommand{\norm}[1]{\left\|#1\right\|}

\newcommand{\normtwo}[1]{\norm{#1}_2}
\newcommand{\norminf}[1]{\norm{#1}_\infty}

\providecommand{\expect}[2]{\ensuremath{\ifthenelse{\equal{#1}{}}{\mathbb{E}}{\mathbb{E}_{#1}}\!\left[#2\right]}\xspace}
\providecommand{\prob}[2]{\ensuremath{\ifthenelse{\equal{#1}{}}{\Pr}{\Pr_{#1}}\!\left[#2\right]}\xspace}

\usepackage{bm}

\newcommand{\abs}[1]{\left|{#1}\right|}

\DeclareMathOperator{\poly}{poly}

\DeclareMathOperator{\supp}{supp}

\usepackage[inline]{enumitem}
\usepackage{array}
\newcolumntype{L}[1]{>{\raggedright\let\newline\\\arraybackslash\hspace{0pt}}m{#1}}
\newcolumntype{C}[1]{>{\centering\let\newline\\\arraybackslash\hspace{0pt}}m{#1}}
\newcolumntype{R}[1]{>{\raggedleft\let\newline\\\arraybackslash\hspace{0pt}}m{#1}}

\title{Streaming Algorithms with Large Approximation Factors}

\begin{document}

\author{Yi Li\footnote{School of Physical and Mathematical Sciences, Nanyang Technological University.  \texttt{yili@ntu.edu.sg}}\\
	   \and
	   Honghao Lin \footnote{Computer Science Department, Carnegie Mellon University.  \texttt{honghaol@andrew.cmu.edu}}
	   %
	   \and
	   David P. Woodruff\footnote{Computer Science Department, Carnegie Mellon University. \texttt{dwoodruf@andrew.cmu.edu}}
        \and
        Yuheng Zhang\footnote{Zhiyuan College, Shanghai Jiao Tong University. \texttt{yveh1999@sjtu.edu.cn}}
	   }
	   
\date{\vspace{-5ex}}	   
\maketitle

\begin{abstract}
We initiate a broad study of classical problems in the streaming model with insertions and deletions in the setting where we allow the approximation factor $\alpha$ to be much larger than $1$. Such algorithms can use significantly less memory than the usual setting for which $\alpha = 1+\epsilon$ for an $\epsilon \in (0,1)$, and are motivated by applications to data-driven algorithm design, among other things. We study large approximations for a number of problems in sketching and streaming, assuming that the underlying $n$-dimensional vector has all coordinates bounded by $M$ throughout the data stream:
\begin{enumerate}

\item For the $\ell_p$ norm\footnote{Here the $\ell_p$ norm $\|x\|_p$ of an $n$-dimensional vector $x$ is $\left (\sum_{i=1}^n |x_i|^p \right )^{1/p}$. For $p < 1$ this is not a norm, but we follow common abuse of notation and call it a norm. Note that it is a well-defined quantity for any $p$.}, $0 <  p \le 2$, we show that obtaining a $\poly(n)$-approximation requires the same amount of memory as obtaining an $O(1)$-approximation for any $M = n^{\Theta(1)}$, which holds even for randomly ordered streams or for streams in the bounded deletion model. Our lower bound also holds for a large class of statistical $M$-estimators. We also give a $2$-pass algorithm that uses less space than the best existing $1$-pass algorithm when the entries of the vector are small. 
\item For estimating the $\ell_p$ norm, $p > 2$, we show an upper bound of $O(n^{1-2/p} (\log n \allowbreak \log M)/\alpha^{2})$ bits for an $\alpha$-approximation, and give a matching lower bound, for almost the full range of $\alpha \geq 1$ for linear sketches. 
We use this to design algorithms with large approximation factors for cascaded norms and rectangle $\ell_p$ norms. 

\item For the $\ell_2$-heavy hitters problem, we show that the known lower bound of $\Omega(k \log n\log M)$ bits
for identifying $(1/k)$-heavy hitters holds even if we are allowed to output items that are $1/(\alpha k)$-heavy, for almost the full range of $\alpha$, provided the algorithm succeeds with probability $1-O(1/n)$. We also obtain a lower bound for linear sketches that is tight even for constant probability algorithms. 

\item For estimating the number $\ell_0$ of distinct elements, we give an  $n^{1/t}$-approximation algorithm using $O(t\log \log M)$ bits of space,  as well as a lower bound of $\Omega(t)$ bits, both excluding the storage of random bits, where $n$ is  the dimension of the underlying frequency vector and $M$ is an upper bound on the magnitude of its coordinates. We also show a separation between $1$ and $2$ passes, and a near-optimal $3$-pass algorithm. 

\item For $\alpha$-approximation to the Schatten-$p$ norm, we give near-optimal $\tilde{O}(n^{2-4/p}/\alpha^4)$ multiplicative approximations for every even integer $p$ and every $\alpha \geq 1$, while for $p$ not an even integer we obtain near-optimal approximations once $\alpha = \Omega(n^{1/q-1/p})$, where $q$ is the largest even integer less than $p$. The latter is surprising as it is unknown what the complexity of Schatten-$p$ norm estimation is for constant approximation; we show once the approximation factor is at least $n^{1/q-1/p}$, we can obtain near-optimal sketching bounds. 
\end{enumerate}
\end{abstract}

%
\newpage

\section{Introduction}
The data stream model is an important model for analyzing massive datasets, where the sheer size of the input imposes severe restrictions on the resources available to an algorithm. Such algorithms have only a small amount of memory and can only make a few passes over the data. Given a stream of elements from some universe, the algorithm maintains a short sketch, or summary, of what it has seen. Often such sketches are linear, which has multiple benefits, e.g., (1) the sketches can handle both insertions and deletions of items, and (2) the sketches are mergeable, meaning that given the sketch of a stream $S$ and the sketch of a stream $S'$, the sketch of the concatenation of  streams $S$ and $S'$ is the sum of the two sketches. 

Many streaming algorithms have been developed to study fundamental problems in databases, such as estimating the number $\ell_0$ of distinct elements, which is useful for query optimization and data mining. Among other things, this statistic can be used for selecting a minimum cost query plan \cite{selinger1989access}, the design of databases \cite{finkelstein1988physical}, OLAP \cite{padmanabhan2003multi,shukla1996storage}, data integration \cite{brown2005toward,dasu2002mining}, and data warehousing \cite{acharya1999aqua}. Other important streaming problems include finding the heavy hitters, also known as the top-$k$, most popular items, frequent items, elephants, or iceberg queries. These can be used in association rules and frequent itemsets \cite{as94,hpy00,hid99,son95,toi96}, and for iceberg queries and iceberg datacubes \cite{br99,fsgmu98,HPDW01}. Other important applications include estimating the frequency moments $F_p$ \cite{ams99}, which for $p \geq 1$ correspond to the $p$-th power of the $\ell_p$ norm of the vector of frequencies of items, where the frequency of an item is its number of occurrences in the stream. For $p \geq 2$, $F_p$ indicates the degree of skew of the data, which may determine the selection of algorithms for data partitioning \cite{dewitt1992practical}. The case $p = 2$ is the self-join size, which is useful for algorithms involving joining a relation with itself. The frequency moments of a vector are special cases of the Schatten-$p$ norms of a matrix, and there is a large body of work in the data stream model studying these intriguing norms \cite{li2016tight,li2017embeddings,lnw19,braverman2018matrix,braverman2020schatten}, as well as the related cascaded norms \cite{cormode2005space,andoni2009efficient,jw09,andoni2011streaming}. 

Given that the memory of a data stream algorithm is often significantly sublinear in the size of a stream $\mathcal{S}$, such algorithms are usually both randomized and approximate, and very often come with a guarantee that for a function $f(\mathcal{S})$, the output $X$ of the algorithm satisfies that
$(1-\epsilon) f(\mathcal{S}) \leq X \leq (1+\epsilon) f(\mathcal{S}),$
with probability at least $2/3$ over the coin tosses of the algorithm, where $\epsilon \in (0,1)$ is a  parameter of the algorithm. Here the $2/3$ probability can typically be amplified to $1-\delta$ by repeating the algorithm $O(\log(1/\delta))$ times independently and outputting the median estimate. While a large body of work in the last two decades has resolved the space complexity of many of the aforementioned problems for $\epsilon \in (0,1)$, for certain applications the lower bounds on the space complexity may be too large to be useful. For such applications it is therefore natural to allow for a larger approximation factor $\alpha > 1$, in the hope of obtaining a smaller amount of memory. Namely, one could instead ask for the output $X$ of the streaming algorithm to satisfy $f(\mathcal{S}) \leq X \leq \alpha \cdot f(\mathcal{S})$. This motivates our main question:

\begin{center}
    \emph{What is the space complexity of classical streaming problems when the approximation factor $\alpha$ is allowed to be much larger than $1$?}
\end{center}

Perhaps surprisingly, this question does not seem to be well-understood, and is in fact open for all of the abovementioned problems in a data stream. There are a few related works, such as \cite{chakrabarti2016strong}, which studies large approximation factors for {\it deterministically} estimating the number of distinct elements, $\ell_p$-estimation, entropy estimation, as well as maximum matching size in a graph stream; see also \cite{assadi2017estimating} for large approximation factor lower bounds for randomized algorithms for maximum matching. Other streaming problems where large approximation factors were studied include dynamic time warping \cite{BCKWY19}, maximum $k$-coverage~\cite{IV19} and the $p$-to-$q$ norms \cite{kmw18}. In contrast to \cite{chakrabarti2016strong}, our focus is on tight bounds for randomized algorithms, for which significantly less memory can be achieved than deterministic algorithms, and for a wide range of fundamental problems in the data stream model that do not appear to have been studied before for large approximation factors.

\subsection{Our Results}
A summary of our upper and lower bounds for a number of data stream problems can be found in Tables \ref{tab:results} and \ref{tab:results_2pass}. 

 For estimating the $\ell_p$ norm for $0 < p \le 2$, we show that obtaining a $\poly(n)$-approximation requires the same amount of memory as obtaining an $O(1)$-approximation, under the common assumption that $M = \poly(n)$. Namely, we show an $\Omega(\log n)$ lower bound even with a random oracle for these problems. Previously, only an $\Omega(1)$ lower bound was known for $\poly(n)$-approximation in this setting. Our result also holds if the stream is randomly ordered, or in the bounded deletion model \cite{jw18}, in which deletions are allowed but the norm should not drop by more than a constant factor from what it was at a previous point in time. Our lower bound can also be extended to a wide class of statistical $M$-estimators. 
 We also show a two-pass algorithm that uses less space than the best existing one-pass algorithm. 
 
 For estimating the $\ell_p$ norm of an underlying $n$-dimensional vector, $p > 2$, we show an upper bound of $O(n^{1-2/p} (\log n \log M)/\alpha^2)$ bits for $\alpha$-approximation for any $\alpha > 1$, and a matching lower bound for almost the full range of $\alpha$ on the bit complexity of linear sketches, which gives a matching streaming lower bound under the conditions of \cite{lnw14}, though these conditions can be restrictive, see, e.g., \cite{KP20} for discussion. One important motivation for studying such norms is to data-augmented streaming algorithms. For example, it was shown in \cite{jiang2019learning} that for estimating the $\ell_p$ norm with a so-called learned oracle, one can achieve an $O(1)$-approximation using $\tilde{O}(n^{1/2 -1/p})$ bits of space. However, this requires a successfully trained oracle, which could have an arbitrarily bad approximation in the worst case. By instead running our worst-case $\tilde{O}(n^{1/4-1/(2p)})$-approximation algorithm for $\ell_p$ estimation with  $\tilde{O}(n^{1/2-1/p})$ bits of memory in parallel, we can ensure that we do at least as well as the learned algorithm in the same amount of memory (up to a constant factor), but we can ensure we never return worse than an $\tilde{O}(n^{1/4-1/(2p)})$-approximation. Another important consequence of our $\ell_p$-estimation algorithm is that it can be used as a subroutine to obtain large approximations for the $(p, q)$-cascaded-norm ($p \ge 1, q > 2$) and rectangle $\ell_p$ ($p > 2$) problems, showing that the previous space bounds can be reduced by an $\alpha^2$ factor for an $\alpha$-approximation. These results are shown in Sections~\ref{sec:cascaded} and~\ref{sec:rectangle}.
 
In the $\ell_2$-heavy hitters problem, the goal is to output a subset $S$ of $\{1, 2, \ldots, n\}$ which contains every $i$ for which $x_i^2 \geq \frac{1}{k} \|x\|_2^2$, and no $i$ for which $x_i^2 < \frac{1}{2k} \|x\|_2^2$. It is known \cite{BIPW10,jst11} that the space complexity of this problem is $\Theta(k \log n \log M)$ bits, if we are promised that $x \in \{-M, \ldots, M\}^n$. A natural relaxation would be to instead require only that $S$ contains every index $i$ for which $x_i^2 \geq \frac{1}{k} \|x\|_2^2$ and no $i$ for which $x_i^2 < \frac{1}{\alpha k} \|x\|_2^2$. We show a strong negative result, that for any $\alpha = O((n/k) (\log \log n)^2/(\log n)^2),$ this problem still requires $\Omega(k \log n \log M)$ bits of memory for any linear sketch, which gives a matching streaming lower bound under the conditions of \cite{lnw14}. For our bit complexity lower bound we assume the algorithm succeeds with probability $1-O(1/n)$, while our sketching dimension lower bound only requires that the algorithm succeeds with constant probability. Interestingly, the proofs of our lower bounds do not use the usual hard instances for finding  $\ell_2$-heavy hitters \cite{BIPW10,jst11}, and instead use a hard instance for $\ell_p$-estimation in \cite{WW15}.

For estimating the number $\ell_0$ of distinct elements, we show that to obtain an $\alpha = n^{1/t}$-approximation, an upper bound of $O(t\log \log M)$ bits is possible and there is a lower bound of $\Omega(t)$ bits, where $n$ denotes the dimension of the underlying frequency vector and $M$ is an upper bound on the absolute value of its coordinates. We state our results in the random oracle model, where a public random string is known to the algorithm. Without such a random string, a simple reduction from the Equality communication problem gives an $\Omega(\log n)$ bit lower bound for any multiplicative approximation, see, e.g., \cite{ams99} for similar arguments\footnote{Briefly, Alice has $x \in \{0,1\}^n$ and inserts $i$ for which $x_i = 1$. Bob has $y \in \{0,1\}^n$ and deletes $i$ for which $y_i = 1$. If $x = y$ then $\ell_0 = 0$, otherwise it is non-zero, and the private coin randomized communication complexity of Equality is $\Omega(\log n)$ bits.}.  Nevertheless, our results are still interesting outside of the random oracle model, since in the common setting of $M \leq \poly(n),$ setting $t = (\log n)/\log \log n$ gives us an $O(\log n)$-approximation with $O(\log n)$ bits of memory, and since $O(\log n)$ bits of randomness is also sufficient, this matches the $\Omega(\log n)$ bit lower bound from the Equality problem. The previous best algorithm \cite{knw10b} required at least $O(\log n \log \log M)$ bits for any multiplicative approximation factor. 
 We also study estimating the number of distinct elements in two and three passes, showing a separation for the problem between one and two passes and a near-optimal three-pass algorithm.


The Schatten-$p$ norm of an $n \times n$ input matrix $A$ is just the $\ell_p$-norm of the vector of singular values of $A$. For $\alpha$-approximation to the Schatten-$p$ norm, we give a linear sketch of dimension $\widetilde{O}(n^{2-4/p}/\alpha^4)$, which is optimal up to logarithmic factors, for every even integer $p$ and every $\alpha \geq 1$, while for $p$ not an even integer we obtain a near-optimal sketch dimension of $\widetilde{O}(n^{2-4/p}/\alpha^4)$ once $\alpha = \Omega(n^{1/q-1/p})$, where $q$ is the largest even integer less than $p$. Interestingly, we obtain the first near-optimal multiplicative approximations for Schatten-$p$ norms for non-integer $p$ for a wide range of non-trivial approximation factors $\alpha$, whereas it is still unknown and a major open question (see, e.g., \cite{lnw19} for discussion) to obtain optimal multiplicative approximations for Schatten-$p$ norms for non-integer $p$ when $\alpha = O(1)$. Our work highlights that surprisingly, the difficulty of this problem stems from the approximation factor rather than the problem being hard for every approximation factor. 

\begin{table}[htbp]
\centering

{
\small
\begin{tabular}{|c|c|c@{\hskip0pt}c|c@{\hskip0pt}l@{\hskip5pt}|}
	\hline
	Problem & \multicolumn{3}{c|}{Large Approx.\ Ratio}   & \multicolumn{2}{c|}{Constant Approx.\ Ratio}\\
	
	\hline	
	\multirow{2}{*}{$\ell_p$ Estimation $(0 < p \le 2)$} & \multirow{2}{*}{$\poly(n)$} & \multirow{2}{*}{$\Omega(\log n)$} & \multirow{2}{*}{Thm~\ref{thm:lb_M}}  & $O(\log n)$ & \cite{knw10}  \\
	& &  & & $\Omega(\log n)$ & \cite{knw10}\\
	\hline
	\multirow{2}{*}{$\ell_p$ Estimation $(p > 2)$} & \multirow{2}{*}{$\alpha$} & $\tilde{O}(n^{1-2/p}/\alpha^2)$ & Thm~\ref{thm:ub_fp} & $\tilde{O}(n^{1-2/p})$ & e.g.~\cite{andoni2011streaming} \\
	& & $\tilde{\Omega}(n^{1-2/p}/\alpha^2)$ &Thm~\ref{thm:lb_fp}& $\tilde{\Omega}(n^{1-2/p})$ &e.g.~\cite{WW15}\\
	
	\hline	
	\multirow{2}{*}{$\ell_2$ Heavy Hitters} & \multirow{2}{*}{$\tilde{O}(n/k)$} & \multirow{2}{*}{$\Omega(k\log^2 n)$} & \multirow{2}{*}{Thm~\ref{thm:hh}} & $O(k\log^2 n)$ &  \\
	& &  & & $\Omega(k\log^2n)$ & \cite{jst11} \\
	\hline
	{$\ell_2$ Heavy Hitters} & \multirow{2}{*}{$\tilde{O}(n/k)$} & \multirow{2}{*}{$\Omega(k\log n)$} & \multirow{2}{*}{Thm~\ref{thm:hh_dimension}} & $O(k\log n)$ &  \\
	(Sketching Dimension)& &  & & $\Omega(k\log n)$ & \cite{PW11} \\
	\hline
	\multirow{2}{*}{Distinct Elements} & \multirow{2}{*}{$n^{1/t}$} & $O(t\log \log n)$ & Thm~\ref{thm:n^1/t-approximator} & $O(\log n \log \log n)$ & \cite{knw10b}   \\
	& & $\Omega(t)$ & Thm~\ref{thm:l0_lb} & $\Omega(\log n \log \log n)$ & \cite{wy19}\\
	\hline
	\multirow{2}{*}{Schatten-$p$ Norm} & \multirow{2}{*}{$\alpha$} & $\tilde{O}(n^{2-4/p}/\alpha^4)$ & Thm~\ref{thm:ub_schatten_even},~\ref{thm:ub_schatten_not_even}  & $O(n^{2-4/p})$ even $p$& \cite{lnw19}  \\
	& & $\Omega(n^{2-4/p}/\alpha^4)$ & Thm~\ref{thm:lb_schatten}& $\Omega(n^{2-4/p})$ & \cite{lnw19}\\
	\hline
	Cascaded Norm & \multirow{2}{*}{$\alpha$} & $\tilde{O}(n^{1-2/p}d^{1-2/q}/\alpha^2)$ & Thm~\ref{thm:ub_cascaded} & $\tilde{O}(n^{1-2/p}d^{1-2/q})$ & \cite{andoni2011streaming} \\
	($p,q > 2$)& & $\Omega(n^{1-2/p}d^{1-2/q}/\alpha^2)$ &Thm~\ref{thm:lb_cascaded}& $\Omega(n^{1-2/p}d^{1-2/q})$ & \cite{jw09}\\
	\hline
	Cascaded Norm & \multirow{2}{*}{$\alpha$} & $\tilde{O}(d^{1-2/q}/\alpha^2)$ & Thm~\ref{thm:ub_cascaded} & $\tilde{O}(d^{1-2/q})$  & \cite{andoni2011streaming} \\
	($1 \le p < 2, q > 2$)& & $\tilde{\Omega}(d^{1-2/q}/\alpha^2)$ &Thm~\ref{thm:lb_cascaded}& $\Omega(d^{1-2/q})$ & \cite{jw09}\\
	\hline
	 Rectangle $F_p$& \multirow{2}{*}{$\alpha$} & $O^\ast(n^{d(1-2/p)} / \alpha^{2})$  & Thm~\ref{thm:rectangle} & $O^\ast(n^{d(1-2/p)})$ & \cite{tw12} \\
	Estimation $(p>2)$& & $\Omega(n^{d(1-2/p)} / \alpha^{2})$ &Thm~\ref{thm:lb_fp}& $\Omega(n^{d(1-2/p)})$ & \cite{tw12}\\
	\hline
\end{tabular}
}
\caption{Summary of previous results and the results obtained in this work. We assume that $M=\poly(n)$ and $p$ is constant. The reported space bounds are measured in bits except for the Schatten-$p$ norm, where we consider the sketching dimension. In this table, $\tilde{\Omega}(f)$ denotes $\Omega(f \poly \log f)$ and, for the rectangle $\ell_p$ estimation problem, $O^\ast(f)$ denotes  $f\cdot \poly(d,\log(mn/\delta))$. For the $\ell_2$-heavy hitters problem, both our lower bound and our upper bound for bit complexity assume that the success probability is at least $1 - O(1/n)$, while for the sketching dimension results we assume constant success probability. 
}\label{tab:results}

{\small
\begin{tabular}{|c|c|c|c|}
	\hline
	Problem & Type& New  Alg & Previous 1-pass Alg\\
	\hline
	\multirow{4}{*}{Distinct Elements} & \multirow{2}{*}{2-pass} & $O(\log n + \eps^{-2}\log \log M (\log (1/\eps) + \log\log M))$ & $O(\eps^{-2} \log n \log \log nM)$\\
	& & Theorem~\ref{thm:two-pass} & \cite{knw10b}\\
    \hhline{~|--|~}
	 & \multirow{2}{*}{3-pass} & $O(\log n + \eps^{-2} (\log (1/\eps) + \log\log M))$ & $\Omega(\eps^{-2} \log n \log \log nM)$\\
	& & Theorem~\ref{thm:three-pass} & \cite{wy19}\\
	\hline
	\multirow{2}{*}{$\ell_p$ Moment $(p \le 2)$} & \multirow{2}{*}{2-pass} & $O(\log n + \eps^{-2} (\log M + \log 1/\eps))$ & $O(\eps^{-2} \log nM)$\\
	& & Theorem~\ref{thm:2pass_lp} &\cite{knw10}\\
	\hline
\end{tabular}
}
\caption{Summary of previous results and the results obtained in this work to obtain a $(1 \pm \eps)$-approximation. The reported space bounds are measured in bit complexity.
}\label{tab:results_2pass}

\end{table}

\subsection{Our Techniques}
For our lower bound for estimating $\ell_p$-norms for $0 < p \leq 2$ (or more generally for $M$-estimators), we give a reduction from the the coin problem introduced in~\cite{braverman2020coin} and strengthened in~\cite{BGZ21}. Consider a sequence of independent coin flips with either a heads probability of $1/2+\beta$ or a heads probability of $1/2-\beta$. The coin problem asks us to distinguish between the two cases with the fewest number of flips. Given a sequence of $n$ coin flips, for an underlying vector $x$ in a stream we can perform $x_1\gets x_1 + 1$ or $x_1\gets x_1 - 1$, depending on whether the coin is a heads or a tail. To ensure a bounded deletion stream, we initialize $x=(2n\beta,0,\dots,0)$. Then, with constant probability, we have $x_1 = 4n\beta \pm O(\sqrt{n})$  in one case and $x_1 = \pm O(\sqrt{n})$ in the other, resulting in an $\alpha$-factor difference in the $\ell_p$-norm when $4n\beta=\Omega(\alpha\sqrt{n})$. Note that our goal is to obtain a lower bound for $\alpha=\omega(1)$. The earlier lower bound for the coin problem~\cite{braverman2020coin} instead considers $\beta \sim 1/\sqrt{n}$, which only translates into $\alpha = \Theta(1)$ at best, for which we know an upper bound of $O(\log n)$ words exists. The newer result~\cite{BGZ21} shows an $O(\log n)$ bit lower bound for $\beta < n^{1/3-\eps}$. Such a $\beta$ translates into $\alpha = n^{\Omega(1)}$, as desired. This is also the first application of the newer result~\cite{BGZ21} to data streams. 

For our upper bound for estimating $\ell_p$-norms for $p > 2$, we connect the problem to an instance of the same problem with a different parameter. Namely, suppose that $q$ is such that $n^{1 - 2/q} = \Theta(n^{1-2/p}/\alpha^2)$, where $\alpha$ is the approximation factor. Then from relationships between norms we have $\norm{x}_p \le \norm{x}_q \le \alpha \norm{x}_p$. Hence, a constant factor approximation to the $\ell_q$ norm actually gives an $\alpha$ approximation to the $\ell_p$ norm. This ``self-reduction" from an instance of the problem under one norm to an instance of the same problem under a different norm also helps us derive our algorithm for estimating the Schatten-$p$ norms of a matrix when $\alpha = \Omega(n^{1/q - 1/p})$, where $q$ is the largest even integer less than $p$. For our lower bound for  $\ell_p$-norm estimation for $p > 2$, we consider the multiparty disjointness ($\mathsf{DISJ}_s^n$) problem in the public-coin simultaneous message passing model, which was initially proposed in~\cite{WW15}. We show that the hard instance can still give a matching lower bound for $\alpha$-approximation if we set the number of players appropriately.

For the $\ell_2$ heavy hitters problem, the usual hard instances for this problem (see, e.g.,~\cite{jst11} and~\cite{BIPW10}) fail to give an  extra $\log n$ factor for large approximations. The reason is that when reducing from the so-called Augmented Indexing problem, to make the two cases distinguishable for an $\alpha$-approximation, one would need to partition the vector into $\log_{\alpha}(n)$ levels, which for $\alpha = n^{\Omega(1)}$, is only $O(1)$. In contrast, we consider the same multiparty disjointess problem we use for the $\ell_p$ norm estimation problem and show that a similar hard instance gives a matching lower bound for the $\ell_2$ heavy hitters problem with a large approximation factor. Thus, we use a fundamentally different hard instance for this problem. 

For our upper bound for estimating the number $\ell_0$ of distinct elements, suppose that the approximation factor $\alpha = n^{1/t}$. We sub-sample the input coordinates into $t$ levels, with a geometrically decreasing sampling probability. In each level, the surviving coordinates are hashed into a constant number of buckets. If the $\ell_0$ of the sub-sampled vector in a level is at most a constant, then only a small number of these buckets will be occupied. Based on this, we can find the specific level $j^*$ for which the $\ell_0$ norm in this level is between $0$ and $n^{1/t}$ and show that after rescaling it is a good estimator to the overall $\ell_0$ of the original vector. To use less memory in each bucket, we choose a random prime $p = \poly(\log M)$ and only store each counter mod $p$. Our lower bound is based on a reduction from the Augmented Indexing problem mentioned above, which in more detail is a two player communication problem in which Alice holds a binary vector $u \in \{0,1\}^t$ and asks for Bob to recover $u_i$ given $u_{i + 1}, \ldots, u_{t}$. We divide the vector $x$ into $l = \Theta(t)$ segments, where the $i$-th segment has length $\Theta(n^{i/l})$, and fill the $i$-th segment with all $1$s if and only if $u_i = 1$. Then $\norm{x}_0$ differs by a factor of $\Theta(n^{1/t})$ between the cases of $u_i = 0$ and $u_i = 1$, whence an $\Omega(t)$ lower bound follows. Despite the fact that a $\log(1/\eps)$-factor gap remains in the upper and lower bounds for $(1 \pm \eps)$-approximation for $\ell_0$ (see, e.g., \cite{2019arXiv190507135D} for discussion), we obtain a tight $\Theta(\log n)$ space bound for $\alpha = \Theta(\log n)$ approximation, for example. Our bounds also show a separation between the estimation of the $\ell_p$-norm ($0 < p \le 2$) and the $\ell_0$-norm with an $n^{\Theta(1)}$-approximation factor, since we show an $\Omega(\log n)$ lower bound via the coin problem for $p > 0$ and $n^{\Omega(1)}$ approximation, while we have an $O(\log \log n)$ upper bound for $p = 0$ and $n^{O(1)}$ approximation. 

We also consider multi-pass algorithms for $\ell_0$ and $\ell_p$ ($0 < p \le 2$) estimation. For the $\ell_0$ estimation problem, we show that if we obtain an $O(\log n)$-approximation in the first pass, then we can obtain a $(1 \pm \eps)$-approximation in the second pass using $O(\eps^{-2} \log \log M(\log(1/\eps) + \log\log M))$ bits of space. This can be further reduced to $O(\eps^{-2}(\log(1/\eps) +\log \log M))$ bits of space using a third pass. For $\ell_p$ estimation, we show that if we can obtain a constant approximation $Z$ in the first pass, then in the second pass, we can sample the coordinates with probability $O(\eps^{-2} M^p / Z)$. Hence, we only need to generate certain $p$-stable random variables used in our algorithm with precision $(M/\eps)^{O(1)}$, from which we obtain an  $O(\eps^{-2}(\log M + \log(1/\eps)))$ bits of space algorithm in the second pass, which is better than the previous result of $O(\eps^{-2}\log nM)$ when $M$ is small. 

\section{Preliminaries}\label{sec:prelim}

\subparagraph{Notation} For a vector $x\in \R^n$, its $\ell_p$ norm is $\norm{x}_p = (\sum_{i=1}^n |x_i|^p)^{\frac{1}{p}}$, where $p\geq 1$. We also write $F_p(x) = \norm{x}_p^p$. We also define $\norm{x}_\infty = \max_i |x_i|$. When $p < 1$, the quantity $\norm{x}_p$ is not a norm though it is a well-defined quantity and $\norm{x}_p^p$ tends to the number of nonzero entries of $x$ as $p\to 0^+$. In view of this limit, we denote the number of nonzero entries of $x$ by $\norm{x}_0$ and also refer to it as $\ell_0$.

For a matrix $A\in \R^{m\times n}$, we define its Schatten-$p$ norm to be $\norm{A}_p = (\sum_{i=1}^{\min\{m,n\}} (\sigma_i(A))^p)^{\frac{1}{p}}$ for each $p\geq 1$, where $\sigma_1(A) \geq \sigma_2(A) \geq \cdots$ are the singular values of $A$. We also define the ($p,q$)-cascaded norm of $A$ to be $\norm{A}_{p,q} \!=\! (\sum_i(\sum_j \! |A_{i,j}|^q)^{\frac{p}{q}})^{\frac{1}{p}}$ for $p,q\geq 1$.

\subparagraph{Turnstile streaming model} In the turnstile model of data streams, there is an underlying $n$-dimensional vector $x$ which is initialized to $0$ and keeps receiving updates of the form $(i,\Delta)\in [n]\times \R$, which represents $x_i\gets x_i + \Delta$. Here $\Delta$ can be either positive or negative. In this paper we assume that the underlying vector is guaranteed to be bounded by $M$, i.e., it always holds that $\norm{x}_\infty\leq M$ throughout the data stream. The length of the stream is denoted by $m$. When the vector $x$ is given by a stream $\cS$ in the turnstile model, we abuse notation and also write $\ell_p(\cS)$ for $\norm{x}_p$.

When the input describes a matrix $A\in \R^{m\times n}$, we can view the matrix as an $mn$-dimensional vector and each item in the stream updates an entry of the matrix.

A variant of the streaming model for a matrix $A$ concerns rectangular updates. Here $x$ is a tensor indexed by $[n]^d$ and each update has the form $(R, \Delta)$, where $R\subseteq [n]^d$ is a rectangle, representing the update $x_i\gets x_i + \Delta$ for all $i\in R$. The rectangle $\ell_p$ problem is considered under this model (see, e.g., \cite{tw12}), which asks to estimate $\norm{A}_p = (\sum_{i\in [n]^d} |x_i|^p)^{1/p}$.

\subparagraph{Subspace Embeddings} Suppose that $A\in\R^{n\times d}$. A matrix $S\in\R^{m\times n}$ is called an $(\eps,\delta)$-subspace-embedding for $A$ if it holds with probability at least $1-\delta$ that $(1-\eps)\norm{Ax}_2\leq \norm{SAx}_2\leq (1+\eps)\norm{Ax}_2$ for all $x\in\R^d$ simultaneously. A classical construction is to take $S$ to be a Gaussian random matrix of i.i.d.\ $N(0,1/m)$ entries, where $m=O((d+\log(1/\delta)/\eps^2)$. Recall the minimax characterization of singular values of a matrix $A$: $\sigma_i(A) = \sup_H \inf_{x\in H: \norm{x}_2=1} \norm{Ax}_2$, where the supremum is taken over all subspaces $H$ such that $\dim(H) = i$. This implies (see e.g., Lemma 7.2 of \cite{lnw19}) that $(1-\eps)\sigma_i(A)\leq \sigma_i(SA)\leq (1+\eps)\sigma_i(A)$ with probability at least $1-\delta$, for all $i=1,\dots,\min\{m,n\}$, i.e., $S$ preserves all singular values of $A$ if $S$ is an $(\eps,\delta)$-subspace-embedding for $A$. 
\section{Lower Bound for $M$-Estimators}
\label{sec:lb_M}
We start by giving a very general lower bound for $M$-estimator estimation with a large approximation factor. $M$-estimators can be seen as generalizations of the $p$-th frequency moments of the underlying vector $x$. We first show this lower bound in the turnstile streaming model and later we will show that it still holds even in the bounded deletion and random order models.

\begin{definition}[$M$-estimator with parameter $\gamma$]
Suppose $G:\R \rightarrow \R_{\ge 0}$ is a function. We say $\norm{y}_G = \sum_i G(y_i)$ is an  $M$-estimator with parameter $\gamma$ if $G$ satisfies the following conditions:
\begin{itemize}
    \item $G(0) = 0$;
    \item $G(x) = G(-x)$;
    \item $G(x)$ is non-decreasing in $|x|$;
    \item For all $x$, $y$ with $|y| > |x| > 0$, $\frac{G(y)}{G(x)} \ge \left|\frac{y}{x}\right|^\gamma$.
\end{itemize}
\end{definition}

We will give a reduction from the following coin problem.
In~\cite{BGZ21}, the authors show an $\Omega(\log n)$ lower bound even when the parameter $\beta$ is allowed to be very small:
\begin{definition}[Coin Problem]
Let $X_1, ..., X_n$ be a stream of i.i.d.\ random bits, which either (1) comes from a distribution with heads probability $\frac{1}{2} + \beta$ or (2) comes from a distribution with heads probability $\frac{1}{2} - \beta$. We are asked to distinguish these two cases at the end of the stream, with probability $2/3$.
\end{definition}

\begin{theorem}[\cite{BGZ21}]
\label{thm:coin}
For all constant $\eps > 0$, any length-$n$ Read-Once Branching Program that solves the coin problem for bias $\beta < n^{-1/3 - \eps}$, requires $n^{\Omega(\eps)}$ width.
\end{theorem}

\begin{corollary}
\label{cor:coin}
For all constants $\eps > 0$, any randomized streaming algorithm that solves the coin problem with bias $\beta < n^{-1/3-\eps}$ requires $\Omega(\log n)$ space. This holds even if we give the algorithm access to an arbitrarily long random tape. 
\end{corollary}

Suppose that we are given an $M$-estimator with parameter $\gamma$. We define a distribution $\cD$ on the sequences of $n$ random bits as follows: suppose we have a sequence $X_1,...,X_n$, which comes from a distribution with heads probability $\frac{1}{2}$ or a distribution with heads probability $\frac{1}{2} + \beta$, where $\beta = n^{-1/3-\eps}$ with a small constant $\eps$. Let $x$ be the underlying vector in the streaming algorithm. During the stream we perform updates $x_1 \gets x_1 + 1$ if $X_i$ = 1, or $x_1 \gets x_1 - 1$  otherwise. We will show that any streaming algorithm that gives an $O(n^{(1/6-\eps)\gamma})$-approximation for $\norm{x}_G$ can distinguish the above two distributions with large constant probability. 
We first analyze the sum $|\sum_{i = 1}^n X_i|$ for these two distributions. 
\begin{lemma}
\label{lem:1/2}
Suppose that the sequence $(X_1, ..., X_n) \in \{\pm 1 \}^n$ comes from the distribution with heads probability $\frac{1}{2}$. Then with probability at least $1 - 1/(4k)$, we have
$\left|\sum_{i = 1}^n X_i\right| \le \sqrt{kn}.$
\end{lemma}
\begin{proof}
We have
$\E \left[\sum_{i = 1}^n X_i\right] = 0$,
and
$
\mathbf{Var}\left[\sum_{i = 1}^n X_i\right] = \sum_{i = 1}^n \mathbf{Var} [X_i] = \frac{n}{4}$. 
From Chebyshev's inequality, we have
$\mathbf{Pr}\left[\left|\sum_{i = 1}^n X_i\right| \le \sqrt{kn}\right] \le \frac{n}{4kn} = \frac{1}{4k}$. 
\end{proof}

\begin{lemma}
\label{lem:1/2+beta}
Suppose that the sequence $(X_1, ..., X_n) \in \{\pm 1 \}^n$ comes from the distribution with heads probability $\frac{1}{2} + n^{-1/3 - \eps}$. Then with probability at least $1 - 1/(4k)$, we have
$\sum_{i = 1}^n X_i \ge 2 n^{2/3-\eps}- \sqrt{kn}$. 
\end{lemma}
\begin{proof}
We have
$\E \left[\sum_{i = 1}^n X_i\right] = 2 n ^{2/3-\eps},$
and
$\mathbf{Var}\left[\sum_{i = 1}^n X_i\right] = \sum_{i = 1}^n \mathbf{Var} [X_i] < \frac{n}{4}$. 
From Chebyshev's inequality, we have
$\mathbf{Pr}\left[\sum_{i = 1}^n X_i \ge 2n^{2/3 - \eps} - \sqrt{kn}\right] < \frac{n}{4kn} = \frac{1}{4k}$. 
\end{proof}

We are now ready to give our lower bound.

\begin{theorem}
\label{thm:lb_M}
Suppose that $G$ is an $M$-estimator with parameter $\gamma$. Then any randomized streaming algorithm which outputs an $O(M^{(1/6-\eps)\gamma})$-approximation to $\norm{x}_{G}$ with probability at least $2/3$ requires $\Omega(\log M)$ bits of space, excluding the storage for random bits.
\end{theorem}
\begin{proof}
Suppose that we have a streaming algorithm which  outputs an $O(M^{(1/6-\eps)\gamma})$-approximation to $\norm{x}_G$. We shall show that we can distinguish the two distributions with bias $\beta$ in the coin problem with large constant probability. 

We initialize the vector $x = (0,0, \ldots ,0)$. Suppose we have a stream of bits $X_1, \ldots, X_M$ coming from the distribution with heads probability $\frac{1}{2}$ or with heads probability $\frac{1}{2} + \beta = \frac{1}{2} + M^{-1/3-\eps}$. Then, during the stream, we perform the update $x_1 \gets x_1 + 1$ if $X_i = 1$ and $x_1 \gets x_1 - 1$ otherwise. 

Let $x^{(0)}$ be the underlying vector if the heads probability for the distribution is $\frac{1}{2}$ and $x^{(1)}$ be the underlying vector if the heads probability is $\frac{1}{2} + \beta$. From Lemmas~\ref{lem:1/2} and~\ref{lem:1/2+beta} we have that with probability at least $9/10$, $\abs{x^{(0)}_1} = O(\sqrt{M})$ at the end of the stream, while $\abs{x^{(1)}_1} = \Omega(M^{2/3 - \eps})$ in the second case. It follows from the definition of $\gamma$ that
$\frac{\norm{x^{(1)}}_G}{\norm{x^{(0)}}_G} = \Omega(M^{(1/6-\eps)\gamma})$ for the two cases. This implies that if the streaming algorithm can output an $O(n^{(1/6-\eps)\gamma})$-approximation to $\norm{x}_G$, then we can distinguish the two distributions with bias $\beta$ in the coin problem. From Corollary~\ref{cor:coin}, such a streaming algorithm needs $\Omega(\log M)$ bits of space.
\end{proof}
\begin{corollary}
Any randomized streaming algorithm outputting an $O(M^{1/6-\eps})$-approximation to $\norm{x}_p^p$ requires $\Omega(p \cdot \log M)$ bits of space (excluding the storage for random bits). Moreover, under the assumption that $M = \poly(n)$, any randomized streaming algorithm outputting a $\poly (n)$-approximation to $\norm{x}_p^p$ requires $\Omega(p \cdot \log n)$ bits of space. 
\end{corollary}
\subsection{Lower Bound with Bounded Deletions}
We next show that our $\Omega(\log M)$ lower bound still holds even with the assumption of bounded deletions. In this model, the updates $\Delta_j$ can be positive or negative, but one is promised that the norm $\normtwo{x}$
never drops by more than an $\alpha$-fraction of what it was at any earlier point in the stream, for a constant parameter
$\alpha$.

To analyze the prefix sum $S_k = \sum_{i = 1}^{k} X_i$, we will need the following inequality.

\begin{lemma}[Kolmogorov's inequality]
\label{lem:doob}
Let $X_1,..., X_n$ be independent random variables with $\mathbb{E}[X_i] = 0$ and $\mathbf{Var}[X_i] < + \infty$ for $i = 1, 2,..., n$. Let $S_k = \sum_{i = 1}^{k} X_i$. Then 
$\mathbf{Pr}\left[\max_{1\le k\le n} |S_k| \ge \lambda\right] \le \frac{\mathbf{Var}[S_n]}{\lambda^2} \;.$
\end{lemma}

The following two lemmas follow easily as corollaries.
\begin{lemma}
\label{lem:1/2_bd}
Suppose that the sequence $(X_1, ..., X_n) \in \{\pm 1 \}^n$ comes from the distribution with heads probability $\frac{1}{2}$. Let $S_k = \sum_{i = 1}^k X_i$. Then with probability at least $1 - 1/(4k)$, we have
$\max_{1\le k \le n}\left| S_k \right| \le \sqrt{kn} \; .$
\end{lemma}
\begin{lemma}
\label{lem:1/2+beta_bd}
Suppose that the sequence $(X_1, ..., X_n) \in \{\pm 1 \}^n$ comes from the distribution with heads probability $\frac{1}{2} + \beta = \frac{1}{2} + n^{-1/3 - \eps}$. Let $S_k = \sum_{i = 1}^k X_i$. Then with probability $1 - 1/(4k)$, we have
$\max_{1\le k \le n} |S_k - 2 \beta k| \le  \sqrt{kn} \; .$
\end{lemma}

We are ready to give the following theorem.
\begin{theorem}
\label{thm:lb_M_bd}
Suppose $G$ is an M-estimator with parameter $\gamma$. Then any randomized streaming algorithm outputting an $O(M^{(1/6-\eps)\gamma})$-approximation of $\norm{x}_{G}$ in the bounded deletion model needs $\Omega(\log M)$ bits, excluding the storage for random bits.
\end{theorem}

\begin{proof}

We will use a similar construction for the hard distribution of that in Theorem~\ref{thm:lb_M}. We initialize a vector $x = (c\sqrt{M},0,\ldots,0)$ with a constant $c$. Suppose we have a stream of binary bits $X_1, \ldots , X_M$ coming from the distribution with heads probability $\frac{1}{2}$ or $\frac{1}{2} + \beta = \frac{1}{2} + M^{-1/3-\eps}$. Then, during the stream, if $X_i = 1$, we perform the update $x_1 \gets x_1 + 1$, otherwise we perform the update $x_1 \gets x_1 - 1$.

Let $x^{(0)}$ and $x^{(1)}$ be the underlying vectors in the two cases. From Lemmas~\ref{lem:1/2_bd} and~\ref{lem:1/2+beta_bd} we obtain that, with high constant probability, $\normtwo{x}$ will not drop by more than a constant factor during the stream. 
Note that, with high constant probability, for the first case $\abs{x^{(0)}_1} = O(\sqrt{M})$ at the end of the stream while for the second case $\abs{x^{(1)}_1} = \Omega(M^{2/3 - \eps})$. From the definition of $\gamma$ we have
$\frac{\norm{x^{(1)}}_G}{\norm{x^{(0)}}_G} \ge \Omega(M^{(1/6-\eps)\gamma})$
for the two cases. The same argument in Theorem~\ref{thm:lb_M} shows that the $\log M$ lower bound still holds even in the bounded deletions model.
\end{proof}

\subsection{Lower Bound in Random Order Model}

In the random order model, we assume the updates $\Delta_j$ come in a random order. We note that the updates for the distribution in Theorem~\ref{thm:lb_M} are a sequence of random $\pm 1$ variables. Hence it satisfies the random order assumption automatically, which means we obtain the following theorem. 

\begin{theorem}
\label{thm:lb_M_ro}
Suppose $G$ is an M-estimator with parameter $\gamma$. Then any randomized streaming algorithm which outputs an $O(M^{(1/6-\eps)\gamma})$-approximation to $\norm{x}_{G}$ in the random order model requires $\Omega(\log M)$ bits of space, excluding the storage for its random bits.
\end{theorem}

\section{$\ell_p$ Estimation $p > 2$}
\label{sec:f_p}

In this section, we consider the problem of estimating $\norm{x}_p$ with a large approximation factor when $p > 2$. We present an algorithm that gives an $\alpha$-approximation to $\norm{x}_p$ using $\tilde{O}(n^{1-2/p}/\alpha^{2})$ bits of space. 
We will also give a matching lower bound for this problem.

 \subparagraph{Upper bound.} 
 Suppose that we want an $\alpha$-approximation where $n^{1 - 2/p}/\alpha^2 = \Omega(1)$ (otherwise there is a trivial $\Omega(1)$ lower bound) and let $q$ be the number such that $n^{1 - 2/q} = \Theta(n^{1-2/p}/\alpha^2)$. Then we have $2 \le q < p$. The following lemma shows that $\norm{x}_q$ is an $\alpha$-approximation to $\norm{x}_p$.
 \begin{lemma}
 Suppose that $p\geq q\geq 2$ and $\alpha\geq 1$ satisfies $n^{1 - 2/q} = n^{1-2/p}/\alpha^2 = \Omega(1)$. Then it holds that $\norm{x}_p \le \norm{x}_q \le \alpha \norm{x}_p$.
 \end{lemma}
 \begin{proof}
 From the assumption we know that $\alpha = n^{1/q - 1/p}$. The $\ell_p$ norm is decreasing in $p$, thus
 $
 \norm{x}_q \ge \norm{x}_p
 $.
 By H\"older's inequality, it also holds that
 $
 \norm{x}_q \le n^{1/q - 1/p} \norm{x}_p = \alpha \norm{x}_p
 $ 
 \end{proof}
 
 The preceding lemma shows that we can use any $O(1)$-approximation algorithm for $\ell_q$ to obtain an $\alpha$-approximation to the $\ell_p$ norm. For example, we can use the $O(n^{1-2/q}\log^2n)$-bit algorithm of~\cite{and17}. Our theorem follows immediately.

\begin{theorem}
\label{thm:ub_fp}
Suppose that $p>2$ is a constant. There is an algorithm whose output is $Z$, which satisfies that $\norm{x}_p \leq Z\leq \alpha\norm{x}_p$ with probability at least $0.9$. Furthermore, 
the algorithm 
uses $O(n^{1-2/p}\log n\log M /\alpha^2)$ bits of space.
\end{theorem}

\subparagraph{Application to Data-augmented Algorithm Design}
One important motivation for $\ell_p$ estimation with large approximation is worst-case guarantees for learning-augmented data stream algorithm design. In~\cite{jiang2019learning}, it was shown that given a heavy hitter oracle which can decide, for each input $i$, whether or not $|x_i|\geq n^{-p/2}\norm{x}_p$, one can estimate $\norm{x}_p$ up to a constant factor with probability at least $0.9$ using $O(n^{1/2-1/p}\log n\log M)$ bits of space. In this case, we say the oracle is successful. However, when the oracle is not successful, there is no worst-case guarantee on the quality of approximation. 
An observation here is that when the oracle is not successful, the estimation will be an under-estimate with high probability.
Letting $\alpha=\Theta(n^{1/4-1/(2p)})$ in the preceding theorem, we obtain an $\alpha$-approximation algorithm whose output $Z$ satisfies $\frac{1}{\alpha} \norm{x}_p \le Z \le \norm{x}_p$ with probability at least $0.9$ using the same $O(n^{1/2-1/p}\allowbreak \log n\log M)$ bits of space. Hence we can run our algorithm and the oracle algorithm in parallel and take a maximum. This guarantees an $\alpha$-approximation in $O(n^{1/2-1/p}\log n\log M)$ bits of space with probability at least $0.8$.

\begin{theorem}
Assuming a successful oracle, there is a streaming algorithm which runs in $O(n^{1/2 - 1/p}\log M \log n)$ bits of space, and for which the output $Z$ satisfies $\norm{x}_p \le Z \le 2 \norm{x}_p$. Moreover, even if the oracle is not successful, the output $Z$ always satisfies $\norm{x}_p \le Z \le n^{1/4 - 1/(2p)}\norm{x}_p$. 
\end{theorem}

\subparagraph{Lower Bound} We next show an $\Omega(n^{1-2/p}(\log(M) \log (1/\delta))/\alpha^2)$ lower bound for obtaining an $\alpha$-approximation to $\norm{x}_p$, or, equivalently, an $\Omega(n^{1-2/p}(\log(M) \log (1/\delta))/\alpha^{2/p})$ lower bound for obtaining an $\alpha$-approximation of $F_p(x)$. We first note that it is easy to get an $\Omega(n^{1-2/p}/\alpha^{2/p})$ lower bound from the following $\ell_\infty^k $ communication problem in~\cite{BJKS04}: there are two parties, Alice and Bob, holding vectors $x, y \in \mathbb{Z}^n$ respectively, and their goal is to decide if $\norminf{x - y} \le 1$ or $\norminf{x - y} \ge k$. This problem requires $\Omega(n/k^2)$ bits of communication~\cite{BJKS04}. Let $k = 2^{1/p}\alpha^{1/p} n^{1/p}$. For the case where $\norminf{x - y} \le 1$, we have $\norm{x - y}_p^p\le n$. For the case where $\norminf{x-y} \ge k$, we have $\norm{x-y}_p^p \ge k^p = 2\alpha n$. Suppose there is an algorithm $\mathcal{A}$ which can output a number $Z$ such that $\norm{x}_p \le Z \le \alpha \norm{x}_p$ with probability at least $2/3$. Then Alice can perform the update $x$ to the algorithm $\mathcal{A}$ and send the memory contents of $\mathcal{A}$ to Bob. Bob then performs the update $-y$ to $\mathcal{A}$. From the discussion above, Bob can determine which of the two cases it is with probability at least $2/3$. 


To obtain a stronger lower bound, we consider the following version of multiparty disjointness ($\mathsf{DISJ}_s^n$), coupled with an input distribution, in the public-coin simultaneous message passing model of communication (SMP), as proposed in~\cite{WW15}. In this setting, there are $s$ players, each of whom has a bit string $\bX_i \in \{0, 1\}^n$ ($i \in [s]$) as input. The inputs are generated according to the following distribution $\eta$.
\begin{itemize}
    \item For each $i \in [n], j \in [s]$, set $\bX_{j, i} \sim B(1/s)$ independently at random.
    \item Pick a uniformly random coordinate $I \in [n]$.
    \item Pick a $Z \in \{0, 1\}$. If $Z = 1$, set $\bX_{j, I} = 1$ for all $j \in [s]$. (If $Z = 0$, keep all coordinates as before.)
\end{itemize}
We call the instance of the inputs $\{\bX_i\}_{i\in [s]}$ a ``YES'' instance when $Z=1$, and a ``NO'' instance when $Z=0$.

The players simultaneously send a message $M_i(\bX_i, R)$ to a referee, where $R$ denotes the public coins shared among the players.
The referee then decides, based on $M_1(\bX_1, R)\dots,M_s(\bX_s, R)$ and $R$, whether $\{\bX_i\}_{i\in [s]}$ forms a YES instance or a NO instance. As observed in~\cite{WW15}, if $X \sim \mathrm{Bin}(s, 1/s)$, then $\mathbf{Pr}[X > \ell] \le (e/\ell)^\ell$. Hence, by a union bound for all coordinates $i\in [n]$, it holds in a NO instance, with probability at least $1-1/\poly(n)$, that $\sum_{j=1}^s \bX_{j,i} \leq c\log n/(\log \log n)$ for all $i\in [n]$. On the other hand, in a YES instance it always holds that $\sum_{j=1}^s \bX_{j,I} = s$. Thus, YES and NO instances are distinguishable for $s = \Omega(\log n / \log \log n)$. 

The following is an augmented version of this problem.
\begin{definition}[$\mathsf{Aug}$-$\mathsf{DISJ}(r, s, \delta)$]\label{def:augdisj}
The augmented disjointness problem $\mathsf{Aug}$-$\mathsf{DISJ}(r, s, \delta)$ is the following $s$-party communication problem. The players receive $r$ instances of $\mathsf{DISJ}_s^n$
$
(\bX_1^1, \dots, \bX_s^1),\allowbreak (\bX_1^2, \dots, \bX_s^2),\allowbreak \dots,\allowbreak (\bX_1^r, \dots, \bX_s^r)
$
and the referee, in addition, receives an index $T \in [r]$ which is unknown to the players, along with the last $(r - T)$ inputs $\left\{(\bX_1^t, \dots, \bX_s^t)\right\}_{t = T + 1}^r$. The inputs are generated according to the following distribution:
\begin{enumerate*}[label=(\roman*)]
    \item $T$ is chosen uniformly at random from $[r]$;
    \item $(\bX_1^T, \dots, \bX_s^T) \sim \eta$;
    \item For each $t\neq T$, $(\bX_1^t, \dots, \bX_s^t) \sim \eta_0$ independently, where $\eta_0$ is the conditional distribution of $\eta$ given $Z = 0$.
\end{enumerate*}

At the end of the protocol, the referee should output whether the $T$-th instance $(\bX_1^T, \dots, \bX_s^T)$ is a YES or a NO instance, i.e., the players need to solve $\mathsf{DISJ}_s^n(\bX_1^T, ..., \bX_s^T)$, with probability $1 - \delta$.
\end{definition}


\begin{theorem}[\cite{WW15}]
\label{thm:ww15}
Suppose that $\delta \ge n \cdot 2^{-s}$. Any deterministic protocol that solves  $\mathsf{Aug}$-$\mathsf{DISJ}(r, s, \delta)$  (as defined in Definition~\ref{def:augdisj}) requires $\Omega(r n \min(\log \frac{1}{\delta}, \log s)/s)$ bits of total communication.
\end{theorem}

\noindent \textit{A Reduction to Streaming:}
To lower bound the space complexity of a streaming algorithm we need a way of
relating it to the communication cost of a protocol for this communication problem. In~\cite{WW15}, the authors use a result of~\cite{lnw14}, showing under certain conditions that any streaming algorithm $\mathcal{A}$ which solves the problem $P$ with probability at least $1 - \delta$ can be converted to a ``path-independent" streaming algorithm $\mathcal{B}$ which solves $P$ with probability at least $1-7\delta$, and which uses the same space up to an additive $(\log n + \log\log m + \log 1/\delta)$ factor. The latter then gives a protocol for the $\mathsf{Aug}$-$\mathsf{DISJ}(r, s, \delta)$ problem. Here path-independence means that the output of the algorithm only depends on the initial state and the underlying frequency vector. In other words, the order of the updates of the same frequency vector will not cause different outputs to such an algorithm. We now assume that the algorithm $\mathcal{A}$ we have enjoys this path-independence property. For a more detailed discussion, we refer the readers to Section~5 in~\cite{WW15}.

Suppose there is a path-indepedent $1$-pass streaming algorithm $\mathcal{A}$ which gives an $\alpha$-approximation to $\norm{x}_p^p$ with probability $1 - \delta$. We shall use this to solve the $\mathsf{Aug}$-$\mathsf{DISJ}(r, s, \delta)$ problem for $s = \Theta(\alpha^{1/p}\allowbreak n^{1/p})$ and $r = \log (M/s)$, from which a space lower bound of $\Omega(n^{1-2/p}\allowbreak \log (M)\allowbreak \log (1/\delta) /\alpha^{2/p})$ bits follows if $ M = \Omega((n\alpha)^{1/p + O(1)})$. 

We design the following protocol $\pi$ between the players and the referee. For each $i \in [s]$, player $i$ has $r$ instances $(\bX_i^1, \bX_i^2, \dots, \bX_i^r)$. Player $i$ performs the update $10^{j-1} \cdot \bX_i^j$ to the algorithm $\mathcal{A}$, for each $j \in [r]$, and sends the memory contents of $\mathcal{A}$ to the referee. Under the path-independence assumption, the referee can determine an equivalent frequency vector (i.e., leading to the same state of the algorithm) from each player and then add up the corresponding updates itself. After receiving $T$ and $\left\{(\bX_1^t, \dots, \bX_s^t)\right\}_{t = T + 1}^r$, the referee performs the update $-10^{j-1} \cdot (\sum_{i = 1}^s \bX_i^j)$ to the algorithm $\mathcal{A}$, for each $j \ge T + 1$. Suppose that $\mathcal{A}$ outputs a set $S$. The referee will output YES is $|S| = 1$ and NO if $S = \emptyset$.

Next we analyze correctness of the above protocol $\pi$. We recall that the referee needs to output the answer to the $T$-th instance. For simplicity, we define $\bY^j = \sum_{i = 1}^s \bX_i^j$ for the $j$-th instance. After taking a union bound, for every instance $j$, $\norminf{\bY^j} \leq c\log n / \log \log n$ if it is a NO instance. Also from a Chernoff bound, it is easy to see that $\normtwo{\bY^j}^2 = \Omega(n)$ for all $j$ with probability at least $1 - e^{-\Omega(n)}$.  
Note that the actual underlying vector that $\mathcal{A}$ maintains has the same output as the frequency vector $\bY = \sum_{t = 1}^{T} 10^{t - 1} \bY^{t}$ after the referee performs the updates. We need the following concentration bounds for $\bY$~\cite{WW15}. We note an omission in the proof in that paper and included a corrected one in Appendix~\ref{sec:concentration Y_{-I}}.

\begin{lemma}[\cite{WW15}]
\label{lem:new_bounds_bin}
Let 
$
\sigma_{r} (\norm{\bY_{-I}}_p^p) = (\E[|\norm{\bY_{-I}}_p^p - \E[\norm{\bY_{-I}}_p^p]|^r])^{1/r}.
$
 It holds that
\begin{gather}
\E[\norm{\bY_{-I}}_p^p] \le K_1^p p^p n \cdot 10^{pT}, \label{new_eqn:E Y_{-I}}\\
\sigma_{r} (\norm{\bY_{-I}}_p^p) \le K_2^p p^p \frac{r}{\ln r}\max\{2^p\sqrt{n}, r^p n^{1/r}\} \cdot 10^{pT}, \label{eqn:new_Y_{-I} moment}
\end{gather}
where $r\geq 2$ is arbitrary and $K_1, K_2 > 0$ are absolute constants.
\end{lemma}
Taking $r = 3\ln n$ in \eqref{eqn:new_Y_{-I} moment} gives that
\begin{equation}\label{new_eqn:Y_{-I} final}
\begin{aligned}
&\quad\ \Pr[|\norm{\bY_{-I}}_p^p - \E[\norm{\bY_{-I}}_p^p]| > 0.1n \cdot 10^{pT}] \\
& \le \Pr[|\norm{\bY_{-I}}_p^p - \E[\norm{\bY_{-I}}_p^p]| > 2 \sigma_{\ell} (\norm{\bY_{-I}}_p^p)]\\
&\le 2^{-r} \le 1/n^2.
\end{aligned}
\end{equation}


We condition on all of the events above. 
 We first notice that in all cases, the value $\norminf{x}$ of the underlying vector $x$ the algorithm $\mathcal{A}$ maintains is less than $\left(\sum_{i = 1}^{r - 1} 10^{i - 1} \cdot \frac{\log n}{\log \log n} + 10^{r - 1} \cdot s \right) < 10^r \cdot s = O(M)$ for $r = \log(M / \alpha^{1/p}n^{1/p})$. 
 
 We first consider the case for which the $T$-th instance is a YES instance. In this case, $\bY_{I}^T = s$ and thus,
$
\norm{\bY}_p^p \ge 10^{(T - 1)p} \cdot s = \Omega(10^{(T-1)p} \cdot\alpha n)
$.

Next consider the case in which the $T$-th instance is a NO instance. In this case, we have from \eqref{new_eqn:E Y_{-I}} and \eqref{new_eqn:Y_{-I} final} that
$
\norm{\bY}_p^p = \norm{\bY_{-I}}_p^p + \bY_I^p \le K_p \cdot 10^{pT}n + 10^{pT} (\frac{\log n}{\log \log n})^p \le 1.1 K_p \cdot 10^{pT}n
$, where $K_p$ is a constant that depends only on $p$. 

From the same argument in Section~\ref{sec:hh} we know that if there is an algorithm that can output a $Z$ such that $\norm{x}_p^p \le Z \le K_p^\prime \alpha \norm{x}_p^p$, we can use this algorithm to solve the $\mathsf{Aug}$-$\mathsf{DISJ}(r, s, \delta)$ problem. From Theorem~\ref{thm:ww15}, we obtain the following theorem. 
\begin{theorem}
\label{thm:lb_fp}
Suppose that $p$ is a constant and $M = \Omega((\alpha n)^{1/p + O(1)})$. Then, for $\delta \ge 2^{-\Theta((n\alpha)^{1/p})}$,
any one-pass streaming algorithm which outputs a number $Z$ for which $\norm{x}_p^p \le Z \le \alpha \norm{x}_p^p$ with probability at least $1 - \delta$ requires $\Omega(n^{1-2/p} \log(M) \log(1/\delta)/\alpha^{2/p})$ bits of space. In particular, when $\delta = \Theta(1/n)$, any one-pass streaming algorithm requires $\Omega(n^{1-2/p} \log(M) \log(n)/\alpha^{2/p})$ bits of space.
\end{theorem}

\section{$\ell_2$ Heavy Hitters}
\label{sec:hh}

In the heavy hitters problem, we want to find a set $S \in [n]$ of indices for the underlying vector $x$ such that:
\begin{enumerate}[label=(\roman*)]
    \item $S$ contains every $i$ such that $|x_i|^2 \ge \frac{1}{k} \normtwo{x}^2$;
    \item $S$ does not contain any $i$ such that $|x_i|^2 < \frac{1}{2k} \normtwo{x}^2$.
\end{enumerate}
We call $S$ a $(1/k)$-heavy set of $x$ if $S$ satisfies the above conditions.
Using the classical Count-Sketch, we can solve the above problem in $O(k \log n \log M)$ bits of space with high probability.
\begin{lemma}
There is a randomized one-pass streaming algorithm which can be implemented in $O(k \log n \log M)$ bits of space such that with probability $1 - 1/\poly(n)$, it can output a $\frac{1}{k}$-heavy set $S$ of $x$.
\end{lemma}

In this section, we consider the following relaxation of the heavy hitters problem, where we want to find a set $S$ of indices such that:
\begin{enumerate}[label=(\roman*)]
    \item $S$ contains every $i$ such that $|x_i|^2 \ge \frac{1}{k} \normtwo{x}^2$;
    \item $S$ does not contain any $i$ such that $|x_i|^2 < \frac{1}{\alpha k} \normtwo{x}^2$.
\end{enumerate}
We call such a set $S$ a $(\frac{1}{k},\alpha)$-heavy set of $S$. Our result is negative, where we show that any one-pass streaming algorithm outputting a $(\frac{1}{k}, \alpha)$-heavy set of $x$ with probability at least $1 - 1/n$ still requires $\Omega(k \log n \log M)$ bits of space if $\alpha = O((n/k) (\log \log n)^2/(\log n)^2)$. 

We will consider the augmented disjointness problem $\mathsf{Aug}$-$\mathsf{DISJ}(r, s, \delta)$ defined in Definition~\ref{def:augdisj}. We first recall the definition of this problem.

\paragraph{Definition~\ref{def:augdisj}}($\mathsf{Aug}$-$\mathsf{DISJ}(r, s, \delta)$)
{\em
The augmented disjointness problem $\mathsf{Aug}$-$\mathsf{DISJ}(r, s, \delta)$ is the following $s$-party communication problem. The players receive $r$ instances of $\mathsf{DISJ}_s^n$
\[
(\bX_1^1, \dots, \bX_s^1),(\bX_1^2, \dots, \bX_s^2), \dots ,(\bX_1^r, \dots, \bX_s^r)
\]
and the referee, in addition, receives an index $T \in [r]$ which is unknown to the players, along with the latest $(r - T)$ inputs $$\left\{(\bX_1^t, \dots, \bX_s^t)\right\}_{t = T + 1}^r.$$ The inputs are generated according to the following distribution.
\begin{itemize}
    \item $T$ is chosen uniformly at random from $[r]$;
    \item $(\bX_1^T, \dots, \bX_s^T) \sim \eta$;
    \item For each $t\neq T$, $(\bX_1^t, \dots, \bX_s^t) \sim \eta_0$ independently, where $\eta_0$ is the conditional distribution of $\eta$ given $Z = 0$.
\end{itemize}
At the end of the protocol, the referee should output whether the $T$-th instance $(\bX_1^T, \dots, \bX_s^T)$ is a YES or a NO instance, i.e., the players need to solve $\mathsf{DISJ}_s^n(\bX_1^T, ..., \bX_s^T)$, with probability $1 - \delta$.
}

Suppose that there is a path-indepedent one-pass streaming algorithm $\mathcal{A}$ which can solve the $(\frac{1}{k}, \alpha)$-heavy hitters problem with probability $1 - O(1/n)$, where $\alpha = O(n/k \cdot (\log\log n / \log n)^2)$. Then we can use it to solve the $\mathsf{Aug}$-$\mathsf{DISJ}(r, s, \delta)$ problem for $r = \log (M/n^{1/2})$, $s = \Theta( \sqrt{n/k})$,  $\delta = 1/n$, from which a space lower bound of $\Omega(k \log n \log M)$ bits  follows if $M = \Omega(n^{1/2 + O(1)})$.  

We design the following protocol $\pi$ between the players and referee. For each $i \in [s]$, player $i$ has the $r$ instances $(\bX_i^1, \bX_i^2, \dots, \bX_i^r)$. Player $i$ then performs the update $10^{j-1} \cdot \bX_i^j$ to the algorithm $\mathcal{A}$ for each $j \in [r]$ and sends the memory of $\mathcal{A}$ to the referee. Under the path-independence assumption, the referee can determine an equivalent frequency vector (i.e., leading to the same state of the algorithm) from each player and then add up the corresponding updates. After receiving $T$ and $\left\{(\bX_1^t, \dots, \bX_s^t)\right\}_{t = T + 1}^r$, the referee performs the update $-10^{j-1} \cdot (\sum_{i = 1}^s \bX_i^j)$ to the algorithm $\mathcal{A}$ for each $j \ge T + 1$. Suppose that $\mathcal{A}$ outputs a set $S$. The referee will output YES is $|S| = 1$ and NO if $S = \emptyset$.

Now we analyze the correctness of the above protocol $\pi$. We recall that the referee needs to output the answer to the $T$-th instance. For simplicity, we define $\bY^j = \sum_{i = 1}^s \bX_i^j$ for the $j$-th instance. 
Recall that after taking a union bound, for every instance $j$, $\norminf{\bY^j} \leq c\log n / \log \log n$ if it is a NO instance. Also from a Chernoff bound, it is easy to see that $\normtwo{\bY^j}^2 = \Omega(n)$ for all $j$ with probability at least $1 - e^{-\Omega(n)}$.  
Note that the actual underlying vector that algorithm $\mathcal{A}$ maintains has the same output as the frequency vector $\bY = \sum_{t = 1}^{T} 10^{t - 1} \bY^{t}$ after the referee performs the updates. 
We need the following concentration bounds, which are a special case of Lemma~\ref{lem:new_bounds_bin} with $p=2$.

\begin{lemma}[\cite{WW15}, special case of Claim~6.2]
\label{lem:bounds_bin}
 It holds that
\begin{gather}
\E\left[\normtwo{\bY_{-I}}^2\right] \le K_1 n \cdot 10^{2T}, \label{eqn:E Y_{-I} special}\\
\sigma_{\ell} \left(\normtwo{\bY_{-I}}^2\right) \le K_2 \frac{\ell}{\ln \ell}\max\{4\sqrt{n}, \ell^2 n^{1/\ell}\} \cdot 10^{2T},\quad \forall \ell\geq 2, \label{eqn:Y_{-I} special moment}
\end{gather}
where $K_1, K_2 > 0$ are absolute constants.
\end{lemma}
Taking $\ell = 3\ln n$ in \eqref{eqn:Y_{-I} special moment} gives that
\begin{equation}\label{eqn:Y_{-I} special final}
\begin{aligned}
&\quad\ \mathbf{Pr}\left[\left|\normtwo{\bY_{-I}}^2 - \E\left[\normtwo{\bY_{-I}}^2\right]\right| > 0.1n \cdot 10^{2T}\right] \\
& \le \mathbf{Pr}\left[\left|\normtwo{\bY_{-I}}^2 - \E\left[\normtwo{\bY_{-I}}^2\right]\right| > 2 \sigma_{\ell} \left(\normtwo{\bY_{-I}}^2\right) \right]\\
&\le 2^{-\ell} \le 1/n^2.
\end{aligned}
\end{equation}

Condition on all of the events above occurring. 
 We first notice that in all cases, the value  $\norminf{x}$ of the underlying vector $x$ that algorithm $\mathcal{A}$ maintains is less than $\left(\sum_{i = 1}^{r - 1} 10^{i - 1} \cdot \frac{\log n}{\log \log n} + 10^{r - 1} \cdot \sqrt{n / k}\right) < 10^r \cdot \sqrt{n} = O(M)$ for $r = \log(M / n^{1/2})$. 
 
 We first consider the case in which the $T$-th instance is a YES instance. In this case, $\bY_{I}^T = s$, and thus
\[
\bY_{I} \ge 10^{T - 1} \cdot s.
\]
Meanwhile, for all $j \ne I$, 
\begin{equation}\label{eqn:Y_j UP}
\bY_{j} \le c\sum_{t = 1}^{T}10^{t - 1} \frac{\log n}{\log \log n} < c \cdot 10^T \frac{\log n}{\log \log n} \;.
\end{equation}
It follows from \eqref{eqn:E Y_{-I} special} and~\eqref{eqn:Y_{-I} special final} that 
\[
 \Omega(10^{2T} \cdot n)\le \normtwo{\bY}^2 = \normtwo{\bY_{-I}}^2 + \bY_{I}^2 \le (K_1 + 0.1) n \cdot 10^{2T} +  s^2.
\]
It thus holds that $\bY_I^2 \geq (1/k) \normtwo{\bY}^2$, or equivalently, $s^2/100 \geq s^2/k + (K_1+0.1)n/k$ when $k > 100$ and $s=\Omega(\sqrt{n/k})$. Furthermore, for $j\neq I$, $\bY_j^2 \leq \normtwo{\bY}^2/(\alpha k)$ when $\alpha = O((n/k) (\log \log n/\log n)^2)$. Therefore, our choices of $k$, $s$ and $\alpha$ imply that the set $S = \{I\}$. 

Now we consider the case when the $T$-th instance is a NO instance. In this case, \eqref{eqn:Y_j UP} holds for all $j\in [n]$. Since $
\normtwo{\bY}^2 \ge  \Omega(10^{2T} \cdot n)$, it follows that $\bY_j^2 \leq \normtwo{\bY}^2/(\alpha k)$ for all $j$, provided that $\alpha = O((n/k) (\log \log n/\log n)^2)$. It follows that $S=\emptyset$.

To conclude, we have proved the following theorem.
\begin{theorem}
\label{thm:hh}
Suppose that $k = \Omega(1)$, $\alpha = O\big(\frac{n}{k} (\frac{\log\log n}{\log n})^2\big)$ and $M = \Omega(n^{1/2 + O(1)})$. Then, any one-pass streaming algorithm that solves the $(1/k, \alpha)$-heavy hitters problem with failure probability $O(1/n)$ requires $\Omega(k \log n \log M)$ bits of space, where the algorithm can store any number of random bits.
\end{theorem}

\paragraph{Sketching dimension lower bound.} One limitation of the above theorem is that it requires the algorithm $\mathcal{A}$ to succeed with high probability. Below we show that any algorithm $\mathcal{A}$ using a linear sketch to solve the $(1/k, \alpha)$-heavy hitters problem with constant probability requires the sketching dimension to be $O(k \log (n/k))$ if $\alpha = O(n/(k\log n))$.

We will consider the following communication game in~\cite{PW11}. Let
$\mathcal{F} \subset \{S \subset [n] \mid \abs{S} = k / 2\}$ be a family
of $k$-sparse supports such that:
\begin{itemize}
\item $\abs{S \Delta S'} \geq k$ for $S \neq S' \in \mathcal{F}$,
\item $\Pr_{S \in \mathcal{F}} [i \in S] = k/(2n)$ for all $i \in [n]$, and
\item $\log \abs{\mathcal{F}} = \Omega(k \log (n/k))$.
\end{itemize}

Let $X = \{x \in \{0, \pm 2\sqrt{n/k}\}^n \mid \supp(x) \in \mathcal{F}\}$.  Let
$w \sim \mathcal{N}(0, I_n)$.  Consider the following process. First, Alice chooses $S \in \mathcal{F}$
uniformly at random. Then $x \in X$ is uniformly at random subject to
$\supp(x) = S$, and then $w\sim \mathcal{N}(0, I_n)$. Then, Alice computes $y = Az = A(x + w)$, where $A \in \R^{m \times n}$ is the sketching matrix in $\mathcal{A}$, and Alice sends $y$ to Bob. Then Bob needs to recover $S$ from $y$.

\begin{theorem}[\cite{PW11}]
Suppose that Bob can recover $S$ with probability at least $2/3$. Then $m = \Omega(k \log(n / k))$. 
\end{theorem}

Next we will show that Alice and Bob an use a $(\frac{1}{k}, \alpha)$-heavy hitters algorithm to solve the communication game above if $\alpha = O(n/(k\log n))$. To show correctness, we need the following bounds for $w$.
\begin{lemma}[folklore]
\label{lem:gaussian_norm}
Suppose that $w \sim \mathcal{N}(0, I_n)$. Then with probability $9/10$ we have the following:
\begin{enumerate*}[label=(\roman*)]
    \item 
    $0.9n \le \normtwo{w}^2 \le 1.1n$;
    \item $\norminf{w} \le c\cdot\sqrt{\log n}$.
\end{enumerate*}
\end{lemma}
Condition on the events above. For each $i \in [n]$, we have 
\begin{alignat*}{3}
z_i &\ge 2\sqrt{n / k} - c\sqrt{\log n} \ge 1.9 \sqrt{n/k}, &\quad i \in S,\\
z_i &\le c\sqrt{\log n}, &\quad i\not\in S.
\end{alignat*}
We also have from Lemma~\ref{lem:gaussian_norm} that 
\[
0.9 n < \normtwo{w}^2 <\normtwo{z}^2 \le \normtwo{w}^2 + \normtwo{x}^2 + 4ck\sqrt{\log n}\sqrt{n/k} < 4n \;.
\]
It follows that any $(\frac{1}{k},\alpha)$-heavy set $T$ will exactly be the support set $S$ if $\alpha = O(n/(k\log n))$. The following theorem is immediate.

\begin{theorem}
\label{thm:hh_dimension}
Suppose that $\alpha = O(n/(k\log n))$. Then, any linear sketching algorithm that solves the $(1/k, \alpha)$-heavy hitters problem with constant probability  requires a sketching dimension of $\Omega(k \log(n/k))$.
\end{theorem}

\section{$\ell_0$ Estimation}
\subsection{One-pass Algorithm}\label{sec:L0}
We describe a randomized algorithm which gives an $n^{1/t}$-approximation with constant probability using $O(t\log\log M))$ bits of space, excluding the storage for random bits. We assume that $n^{1/t}\geq c_2$ for some constant $c_2$, otherwise an optimal algorithm is known~\cite{knw10}.

The algorithm is presented in Algorithm~\ref{alg:ell_0}. The idea behind the algorithm is to subsample the coordinates at $t$ levels, with a geometrically decreasing sampling probability. In each level, the surviving coordinates are hashed into a constant number of buckets. If the $\ell_0$ of the subvector (which is the vector of surviving coordinates) at a level is at most a constant, then only a small number of these buckets will be  occupied. Otherwise, all the buckets will be occupied with high probability. Based on this, we design a criterion to determine the occupancy of these buckets to infer the $\ell_0$ of the subvector at a level.
Finally, we find the specific level $j^\ast$ such that the $\ell_0$ in level $j^\ast$ is between $0$ and at most $n^{1/t}$, and then it can be shown that $n^{j^\ast/t}$ is a good estimator to the overall $\ell_0$.

\begin{algorithm}[t]
    \caption{$n^{1/t}$-approximator for $\ell_0$}
    \label{alg:ell_0}
    \SetAlgoLined

    Initialize $cKt$ counters $C_{1, 1, 1}, \dots, C_{t, K, c}$ to $0$\;
    $c_1 \gets \beta \sqrt{c}$\;
    Initialize pairwise independent hash functions $h: [n] \rightarrow [n]$, $g: [n] \rightarrow [c]$\;
    Initialize $K$ $4$-wise independent hash functions $s_i: [n] \rightarrow \{-1, 1\}$\;
    Pick a prime $p \in \Theta(c^3\log^2 M)$\;

    \ForEach{$(x, v)$ in the data stream}{
        $b \gets$ the largest $j$ such that $h(x) \bmod \lfloor n^{1/t} \rfloor^j = 0$\;
        \For{$i \gets 1 $ \KwTo $b$}{
            \For {$j \gets 1$ \KwTo $K$}{
                $C_{i, j, g(x)} \gets (C_{i, j, g(x)} + v \cdot s_j(x)) \bmod p$\;
            }
        }
    }
    
    \uIf{there exists $j$ such that $|\{k \mid \exists l,  C_{j, l, k} \neq 0\}| > c_1$}{
        $J \gets$ the largest $j$ such that $|\{k \mid \exists l,  C_{j, l, k} \neq 0\}| > c_1$\; \label{alg:ell_0:b}
    }
    \Else{
        $J \gets 0$\;
    }
    \KwRet{$c_2 n^{J/t}$}\;
\end{algorithm}

We need the following simple lemma for our analysis. All the omitted proofs in this subsection can be found in Appendix~\ref{sec:l0_proofs}.


\begin{lemma}\label{lem:bucket_value} Suppose that $v_1,\dots,v_t\in \R$ are not all zero and $s_1,\dots,s_t$ are $4$-wise independent Rademacher variables. Let $X = \sum_i v_i s_i$. Then $\Pr\{X = 0\} \leq 2/3$.
\end{lemma}

To show the correctness of the algorithm, let $j^{*}$ be the largest $j$ such that $\E[\ell_0(S^{j})] \geq c_2$, then $c_2 \leq \E[\ell_0(S^{j^\ast})] \leq c_2 n^{1/t}$.

\begin{lemma}\label{lem:b}
With probability at least $0.9$, it holds that $J=j^\ast$ or $J=j^\ast + 1$, where $J$ is as found by the algorithm in Line~\ref{alg:ell_0:b}.
\end{lemma}

\begin{proof}
For a fixed $j > j^{*} + 1$, we have from Markov's inequality and the definition of $j^\ast$ that 
$\Pr[\ell_0(S^j) > c_1] \leq E[\ell_0(S^{j})]/c_1 \leq (c_2/c_1)n^{(j^{*} - j + 1)/t}.$
Let $\cE$ denote the event that $\ell_0(S^j) > c_1$ for some $j>j^\ast + 1$. Then, by a union bound,
$\Pr(\cE) \leq \sum_{j>j^\ast + 1} (c_2/c_1)n^{(j^{*} - j + 1)/t} \leq (c_2/c_1)n^{-1/t}.$

Denote by $\cE'$ the event that $\ell_0(S^{j^\ast}) \leq c_1$. Note that $\mathbf{Var}[\ell_0(S^{j^\ast})] \leq \E[\ell_0(S^{j^\ast})]$. By Chebyshev's inequality, we have that 
\[
\Pr\left[\left|\ell_0(S^{j^\ast}) - \E[\ell_0(S^{j^\ast})]\right| \geq \E[\ell_0(S^{j^\ast})] - c_1\right] 
    \leq \E[\ell_0(S^{j^\ast})]/(\E[\ell_0(S^{j^\ast})] - c_1)^2 \leq c_2/(c_2 - c_1)^2.
\] 
The last inequality is due to the fact that $f(x) = x/(x-c_1)^2$ is decreasing when $x\geq c_1$. Hence,
$\Pr[\cE'] \leq c_2/(c_2 - c_1)^2.$
Next we condition on $\overline{\cE}$ and $\overline{\cE'}$.
For $j > j^\ast + 1$, since we condition on $\overline{\cE}$, it holds that $\ell_0(S^{j}) \leq c_1$ and thus $|\{k \mid \exists l, C_{j, l, k} \neq 0\}| \leq c_1$. For $j^\ast$, since we condition on $\bar{\cE'}$, we have that $\ell_0(S^{j^\ast}) > c_1$. Let $c=c_1^2/\beta^2$ for a $\beta\in(0,1)$ to be determined. With constant probability, there exists a subset of size $c_1$ which is perfectly hashed, and therefore, at least $c_1$ buckets are occupied. Let $\cF$ be the event that $c_1$ coordinates are perfectly hashed.
We have
$\Pr[\cF] = \prod_{i=0}^{c_1-1} \left(1 - i/c\right) \geq \left( 1 - c_1/c\right)^{c_1} \geq 1 - \beta^2.$

Now we bound the probability of the event $\cQ$ that the values of these $c_1$ buckets remain nonzero. For each bucket, we take $K\geq \log_{3/2}(100c_1)$ independent copies of random sign vectors. Whenever there is a nonzero coordinate of $x$ that is hashed into the bucket, it follows from Lemma~\ref{lem:bucket_value} that at least one of the $K$ copies is nonzero with probability at least $1-1/(100c_1)$. However, the bucket values we maintain are taken modulo $p$, and so we also need to bound the probability that $p$ divides the value in the bucket. Since the value in each bucket value is bounded by $M$, it has at most $\log M$ prime factors. Pick a prime $p$ from $[D, D^3]$ uniformly at random, where $D = Kc\log M$ for some constant $K$ to be determined. There are at least $O(c^2 \log^2 M)$ choices for $p$. Adjusting $K$, we can ensure that $p$ divides the bucket value with probability at most $1/(100c_1)$. A union bound over $c_1$ buckets gives that
$\Pr(\cQ) = 1 - c_1\cdot 2/(100c_1) = 0.02.$
Therefore, when conditioned on $\overline{\cE}$ and $\overline{\cE'}$, the algorithm will compute $J=j^\ast$ or $J=j^\ast+1$ except with probability at most $\beta^2 + 0.02$. Removing the conditioning on $\overline{\cE}$ and $\overline{\cE'}$, the overall failure probability is at most
$c_2/(c_1 n^{1/t}) + c_2/(c_2-c_1)^2 + \beta^2 + 0.02 < 1/10$ 
if we choose $c_1 = 25$, $c_2 = 100$ and $\beta = 1/8$.
\end{proof}

Our main theorem is now immediate.

\begin{theorem}\label{thm:n^1/t-approximator}
Algorithm~\ref{alg:ell_0} outputs $Z$, which with probability at least $0.9$ satisfies that $\ell_0/n^{1/t}\leq Z\leq L_0n^{1/t}$. Furthermore, Algorithm~\ref{alg:ell_0} uses $O(t\log\log M)$ bits of space, excluding its random tape.
\end{theorem}
\begin{proof}
Note that $\E[\ell_0(S^{j^\ast})] = \ell_0/n^{j^\ast/t}$ and so $c_2\leq \ell_0/n^{j^\ast/t}\leq c_2n^{1/t}$. It then follows from Lemma~\ref{lem:b} that $\ell_0/(c_2n^{1/t}) \leq n^{J/t}\leq \ell_0 n^{1/t}/c_2$ with probability at least $0.9$. The correctness of the algorithm follows immediately. 

Since the algorithm assumes a random oracle, we do not consider the space for storing the hash functions and the random signs. The space is clearly dominated by the $O(t)$ counters $C_{j,l,k}$. Since $\log p = O(\log\log M)$, each counter takes $O(\log\log M)$ bits of space. Hence, the algorithm uses $O(t\log\log M)$ bits of space in total.
\end{proof}

\begin{remark}
Algorithm~\ref{alg:ell_0} uses $O(\log n)$ random bits since the hash functions $h$, $g$ are pairwise independent and the $s_i$ are $4$-wise independent.
\end{remark}

\subsection{Lower Bound}
We now prove a space lower bound of $\Omega(t)$ bits for estimating $\ell_0$ up to an  $n^{1/t}$-approximation factor. Our lower bound holds even if the algorithm has access to an arbitrarily long random tape, which we do not charge for in its space. We reduce the $\ell_0$ estimation problem to the Augmented Indexing communication problem, in the one-way public coin model, which we now define. We assume that $t= O(\log n)$.

\begin{definition}[Augmented Indexing]
Alice has a string $u \in \{0, 1\}^l$, Bob has an index $i^\ast\in [l]$ and $u_{i^\ast+1}, \dots, u_{l}$. Alice is allowed to send a single message to Bob, and Bob wants to learn $u_{i^\ast}$ from Alice with probability at least $2/3$.
\end{definition}

\begin{lemma}[{\cite{bar2004sketching}}] \label{lem:augindex}
The one-way communication complexity of Augmented Indexing is $\Omega(l)$ in the public coin model.
\end{lemma}

Assume we have a streaming algorithm $\cA$. Alice runs $\cA$ on her stream $s(a)$, then sends the state of $\cA$ to Bob. Bob feeds his stream $s(b)$ into $\cA$ and obtains an estimate of $\ell_0$. We show how to design $s(a)$ and $s(b)$ so that Bob can solve the Augmented Indexing problem.
%

Without loss of generality, we assume that $t$ is divisible by $8$. Let $u$ be the vector in an instance of the Augmented Indexing problem with $l = t/8$. We shall create an input vector $x$ for the $\ell_0$ estimation problem. 
We divide the vector $x$ into $l$ segments, where the $i$-th segment is of length $\frac{1}{2}\lceil n^{i/l}\rceil$. Alice fills in the $i$-th segment with $1$ if $u_i = 1$ by adding a $1$ to each coordinate in the segment. Bob sees $u_{i^\ast+1},\dots,u_l$ and clears the $j$-th segment if $u_j = 1$ ($j>i^\ast$) by adding a $-1$ to each coordinate in the segment. Bob obtains an estimate $Z_1$ to $\norm{x}_0$. Then he adds $1$s to the $i^\ast$-th segment and obtains a new estimate $Z_2$ to $\norm{x}_0$. If $Z_2 \geq n^{4/t}Z_1$, Bob outputs $u_{i^\ast} = 0$, otherwise he outputs $u_{i^\ast} = 1$. A straightforward calculation gives the following lemma.

\begin{lemma}\label{lem:length of x}
The length of $x$ is at most $n$, for $n$ large enough.
\end{lemma}

\begin{theorem}
\label{thm:l0_lb}
Estimating $\ell_0$ with approximation factor  $n^{1/t}$ requires $\Omega(t)$ bits, even if the algorithm has an arbitrarily long random tape. 
\end{theorem}

\begin{proof}
By Lemma~\ref{lem:length of x}, our data stream is well-defined. 
Assume that $\Alg$ provides an $n^{1/t}$-approximation to $\norm{x}_0$ with success probability $p$. 

If $x_i = 1$, the $i^\ast$-th segment has been filled in by Alice, so Bob's last filling has no effect on $\norm{x}_0$. Hence $Z_1$ and $Z_2$ are estimates of the same value and $Z_2/Z_1\leq n^{2/t}$. 

If $x_i = 0$, before Bob's last filling, 
$\norm{x}_0 = \sum_{j=1}^{i^\ast-1} \frac{1}{2}\lceil n^{j/l}\rceil \leq \frac{i^{\ast}}{2} + \frac{1}{2} n^{1/l} \frac{n^{(i^\ast-1)/l} - 1}{n^{1/l} - 1} \leq n^{(i^\ast-1)/l}$, 
when $n$ is large enough. Hence $Z_1\leq n^{1/t}n^{8(i^\ast-1)/t} = n^{(8i^\ast-7)/t}$. After Bob's last filling, $\norm{x}_0\geq \frac{1}{2}n^{i^\ast/l}$ and thus $Z_2\geq \frac{1}{2}n^{(8i^\ast-1)/t}$. Therefore $Z_2/Z_1 \geq \frac{1}{2}n^{6/t}$. 

By our reduction, Bob can output the correct $x_i$ with probability at least $1-2(1-p)$, as $\Alg$ needs to estimate $\ell_0$ correctly twice. We can repeat $O(1)$ times to make this probability at least $2/3$. The lower bound follows from Lemma~\ref{lem:augindex}.
\end{proof}

\subsection{Multi-pass $\ell_0$ Estimation}

\subparagraph{Two-pass algorithm} For $(1 \pm \eps)$-approximation to $\ell_0$ in a turnstile stream, the best one-pass algorithm uses $O(\eps^{-2}\log n (\log (1/\eps) + \log\log M))$ bits of space~\cite{knw10b}, which is still more than the $O(\eps^{-2} + \log n)$ bits of space in insertion-only streams~\cite{knw10b}.
In this section we show that with an additional pass, we can reduce the space complexity to $O(\log n + \eps^{-2} \log\log M(\log (1/\eps) + \log\log M))$ bits. By allowing a third pass, we can further reduce it to $O(\log n + \eps^{-2} (\log (1/\eps) + \log\log M))$ bits of space. This result is nearly optimal in view of an $O(1)$-pass lower bound of $\Omega(\eps^{-2} + \log n)$ bits of space \cite{ams99,CR12}. Combined with the $\Omega(\log n \log \log M)$ $1$-pass lower bound for $O(1)$-approximation in~\cite{wy19}, it also implies a \textit{separation} for $\ell_0$ estimation problem between $1$ and $2$ passes.

Our algorithm is based on the $1$-pass algorithm in~\cite{knw10b}, which we first describe.  
In~\cite{knw10b}, the algorithm first obtains a constant factor approximation and then improves it to a $(1\pm\eps)$-approximation. To obtain a constant factor approximation, it subsamples the data stream into $\log n$ levels and in each level hashes the indices into $O(1/\eps^2)$ buckets using a pairwise independent hash function and maintain their sum modulo a prime $p = O(\eps^{-2}\log M)$. Each sampling level can be maintained in $O(\eps^{-2}(\log (1/\eps) + \log\log M ))$ bits and the overall space complexity is further multiplied by a $\log n$ factor. We observe that finding a $(\log n)$-approximation in the first pass only requires $O(\log n)$ bits of memory, as opposed to the usual $O(\log n \log \log n)$ bits that an $O(1)$-approximation requires. Given an $O(\log n)$-approximation, it then suffices to maintain only $O(\log\log n)$ subsampling levels in the second pass, which reduces the space in both passes. We state our result in the following theorem (the details can be found in Appendix~\ref{sec:L0_multipass}).

\begin{theorem}\label{thm:two-pass}
There exists an absolute constant $\eps_0$ and a two-pass algorithm such that the following holds. For all $\eps\in(0,\eps_0)$, the algorithm outputs $Z$ satisfying $(1-\eps)\ell_0\leq Z\leq (1+\eps)\ell_0$ with probability at least $0.8$. The algorithm uses $O(\log n + \eps^{-2}\log \log M (\log (1/\eps) + \log\log M))$ bits of space.
\end{theorem}

\subparagraph{Three-pass algorithm}
If we are allowed a third pass, we can take $\eps=O(1)$ for the two-pass algorithm and obtain a constant-factor approximation in 
$O(\log n + \log\log^2 M) = O(\log n)$ bits of space. Then we need only maintain one level as in our brief review of the algorithm~\cite{knw10b} to obtain a $(1\pm\eps)$-approximation of $\ell_0$ using another 
$O(\log n + \eps^{-2}(\log (1/\eps) + \log\log M))$ bits. 
\begin{theorem}
\label{thm:three-pass}
There exists an absolute constant $\eps_0$ and a three-pass algorithm such that the following holds. For all $\eps\in(0,\eps_0)$, the algorithm outputs $Z$ satisfying $(1-\eps)\ell_0\leq Z\leq (1+\eps)\ell_0$ with probability at least $0.75$. Furthermore, the algorithm uses $O(\log n + \eps^{-2} (\log (1/\eps) + \log\log M))$ bits of space.
\end{theorem}
\section{Two-Pass Algorithm for $F_p$ ($0 < p \leq 2$)} 
\label{sec:ellp_2}
As we have shown in the previous section, for the $\norm{x}_p^p$ estimation problem, even a large approximation also requires $\Omega(\log n)$ bits of space. In this section, we will show that after obtaining a constant approximation to $\norm{x}_p^p$ in the first pass using $O(\log n)$ bits, we can obtain a $(1 \pm \eps)$-approximation to $\norm{x}_p^p$ using $O(\log n + \eps^{-2}(\log M + \log \frac{1}{\eps}))$ bits. This is better than the previous $O(\eps^{-2}\log nM)$ space bound in one-pass if $M$ is small. Before we give our algorithm, we will first review the algorithm for the $1$-pass case.

\subsection{Review of Previous Algorithm}

\label{sec:ell_p_review}
The method proposed in~\cite{ind06} is based on $p$-stable random variables. 

\begin{definition}[Zolotarev~\cite{zolotarev1986one}]
For $0 < p < 2$, there exists a probability distribution $\mathcal{D}_p$ called the $p$-stable distribution, which satisfies the following property. For any positive integer $n$ and vector $x \in \R_n$, if $Z_1,...,Z_n \sim \mathcal{D}_p$ are independent, then $\sum_{j=1}^n Z_jx_j \sim  \|x\|_pZ$ for $Z \sim \mathcal{D}_p$.
\end{definition}

In~\cite{ind06}, the algorithm takes an $O(\eps^{-2}) \times n$ matrix $A$ whose rows are independent $p$-stable random variable and maintains the matrix-vector product $Ax$ during the stream. Intuitively, each entry of $Ax$ can be seen as a sample from the distribution $\norm{x}_pZ$ for $Z \sim \gD_p$. It was shown that the median of $\{(Ax)_i\}$ is a good approximation to $\norm{x}_p$, after dividing by the approximate median of $|\gD_p|$.

The remaining question is how to generate the independent $p$-stable random variables in very small space. In~\cite{ind06}, the $p$-stable random variables in the same row need to be fully independent and the author used Nisan's Pseudorandom Generator (\cite{Nis92}) to derandomize the algorithm. It was later shown in \cite{knw10} that it suffices for the entries in each row of $A$ to be $O(\eps^{-p})$-wise independent, and the seeds used to generate each row need only be pairwise independent. It was shown in~\cite{cha1976} that the $p$-stable random variable $X$ can be generated by taking $\theta$ uniform in $[-\frac{\pi}{2}, \frac{\pi}{2}]$ and $t$ uniform in $[0, 1]$, and letting
\[
X = f(\theta, t) = \frac{\sin (p\theta)}{\cos^{1/p}(\theta)} \cdot \left(\frac{\cos (\theta(1-p))}{\log(1/t)}\right)^{(1-p)/p} \;.
\]
One can show that with high constant probability, that it suffices to choose $t$ and $\theta$ with precision $(\eps / n)^{O(1)}$. The statements above imply that $O(\eps^{-2} \log n)$ bits of space suffice for $p < 2$ to generate the matrix $A$. Hence, we obtain the following theorem. We refer the reader for a more detailed discussion in~\cite{jel11}.

\begin{theorem}[\cite{knw10}]
\label{thm:knw10}
Suppose that $0 < p < 2$. There is a one-pass streaming algorithm which can be implemented in $O(\eps^{-2}\log (nM) )$ bits of space and which outputs a $(1 \pm \eps)$-approximation to  $\norm{x}_p^p$ with probability at least $9 / 10$.
\end{theorem}

\begin{remark}
The case $p=2$ admits a simpler algorithm. As shown in~\cite{ams99}, we can replace $p$-stable random variables with $4$-wise independent random signs and obtain an algorithm that uses $O(\eps^{-2} \log nM)$ bits of space.
\end{remark}

\subsection{New Two-pass Algorithm}

In this section, we will give our two-pass algorithm.
We begin with the following lemma, which shows that when all entries of $x$ are small, uniform sampling is enough to obtain a good approximation. 

\begin{lemma}
\label{lem:uniform_sample}
Suppose that $\norm{x}_\infty^p \le \norm{x}_p^p/T$. Let $\sigma_i$ be pairwise independent Bernoulli random variables with $\E \sigma_i = q = O(\eps^{-2}/T)$. Then, with probability at least $9 / 10$,
\[
\frac{1}{q}\sum_{i\in [n]} \sigma_i |x_i|^p = (1 \pm \eps) \norm{x}_p^p \;.
\]
\end{lemma}

\begin{proof}
Let $Y = \frac{1}{q} \sum_{i \in [n]} \sigma_i |x_i|^p$. Then we have
\[
    \E \left[Y\right] = \frac{1}{q} \E \left[\sum_{i \in [n]} \sigma_i|x_i|^p \right] = \frac{1}{q}  \sum_{i \in [n]}\E \left[\sigma_i |x_i|^p\right] 
    = \frac{1}{q} \cdot q \cdot \sum_{i \in [n]} |x_i|^p = \norm{x}_p^p \;,
\]
and
\begin{align*}
&\quad\ \mathbf{Var}[Y]  =  \mathbb{E}[Y^2] - {\mathbb{E}[Y]}^{2}  \\
& = \frac{1}{q^2} \left( \mathbb{E}\left[ 
\sum \limits_{i} {\sigma_i {|x_i|}^{2p}}  + \sum \limits_{i \neq j} \sigma_i \sigma_j {|x_i|}^p{|x_j|}^p \right] \right) - \|x\|_p^{2p}
\\
  & \le \frac{1}{q^2} \left( \mathbb{E}
\sum \limits_{i} {\sigma_i {|x_i|}^{2p}}  + \E \sum \limits_{i, j} q^2 {|x_i|}^p{|x_j|}^p \right)  - \|x\|_p^{2p}
  \\
  & =  \frac{1}{q}\sum \limits_{i \in [n]} {|x_i|}^{2p} \\
  & \le \frac{1}{q} \left(\max_{i \in [n]}  |x_i|^p \right) \cdot \left(\sum_{i \in [n]} |x_i|^p\right) \\
  & \le \frac{1}{q} \cdot \frac{\norm{x}_p^p}{T} \norm{x}_p^p\\
  & \le O(\eps^2 \norm{x}_p^{2p})
  \; .
\end{align*}
From Chebyshev’s inequality, we have that 
\[
    \mathbf{Pr} \left[\left|Y - \|x\|_p^p\right| > \eps \|x\|_p^p \right]  \le \frac{1}{10} \;.\qedhere
\]
\end{proof}

Now we are ready to prove our theorem.
\begin{theorem}
\label{thm:2pass_lp}
Suppose that $0 < p \le 2$. There is a two-pass streaming algorithm which can be implemented in $O(\log n + \eps^{-2}(\log M + \log \frac{1}{\eps}))$ bits of space and which outputs a $(1 \pm \eps)$-approximation to $\norm{x}_p^p$ with probability at least $9 / 10$.
\end{theorem}

\begin{proof}
From Theorem~\ref{thm:knw10}, we can obtain a $2$-approximation $Z$ in the first-pass in $O(\log nM) = O(\log n)$ bits of space. It holds that $Z \le \norm{x}_p^p \le 2Z$.

We have for each coordinate $x_i$ that
\[
|x_i|^p \le M^p = \frac{\norm{x}_p^p}{\norm{x}_p^p/M^p} \le \frac{\norm{x}_p^p}{Z/M^p}  \;.
\]
Hence, by Lemma~\ref{lem:uniform_sample}, if we uniformly sample a subset\footnote{From Lemma~\ref{lem:uniform_sample} we know that pairwise independence is enough for our sampling here.} $S \in [n]$ with sampling probability $O(\eps^{-2}M^p / Z)$, with high constant probability we will have
\[
\sum_{i \in S} |x_i|^p = (1 \pm \eps) \norm{x}_p^p \;,
\]
which means we only need to obtain a $(1 \pm \epsilon)$-approximation of $\sum_{i \in S} |x_i|^p$. The expected size of $S$ is $O(n \cdot \eps^{-2}M^p/Z)$ and there are at most $2Z$ non-zero coordinates in $x$. Hence the expected number of non-zero coordinates in $S$ is
\[
O\left(\eps^{-2}n \cdot \frac{M^p}{Z} \cdot \frac{2Z}{n}\right) = O(\eps^{-2} M^p) \;.
\]
It follows from Markov's inequality that $S$ contains $O(\eps^{-2} M^p)$ nonzero coordinates with probability at least $19/20$. 

Condition on this event. From the above, if we run the $p$-stable sketch on $S$, each $p$-stable random variable can be rounded to an additive integer multiple of $(\eps/M)^{O(1)}$, and bounded above by $(M/\eps)^{O(1)}$, which means that $O(\eps^{-2} (\log M + \log(1/\eps)))$ bits is enough to generate the $p$-stable sketching matrix $A$. Recall that there are at most $O(M^p/\eps^2)$ non-zero coordinates in $S$ and each coordinate is at most $M$. Hence each coordinate of the sketch $Ax$ will be an integer of value at most $\poly(M/\eps)$, and we only need  $O(\eps^{-2}(\log M + \log 1/\eps))$ bits of space for storing $Ax$. Putting everything together, we see that the overall space that the algorithm uses is 
\[
O(\log n) + O(\eps^{-2} \log M) + O(\eps^{-2}(\log M + \log 1/\eps)) 
= O(\log n + (\eps^{-2}(\log M + \log 1/\eps))) \; . \qedhere
\] 
\end{proof}
\section{Schatten-$p$ Norm Estimation}

In this section, we consider approximating the Schatten-$p$ norm $\norm{A}_p$ of a given matrix $A$ with large approximation factor $\alpha$, where $\sigma_i(A)$ is the $i$-th singular value of $A$. We assume $A \in \R^{n \times n}$ here because for a general matrix $A \in \R^{n \times d}$, we can first apply a subspace embedding to the left or to the right of $A$ to
preserve each of its singular values up to a constant factor and then pad with zero rows or columns (see, e.g., Appendix C of \cite{lnw14} for the details of this argument). As in the majority of previous work on Schatten norm estimation, we focus on the sketching dimension complexity. 

\paragraph{Upper Bound} We will show that for an even integer $p$ and an arbitrary $\alpha = \Omega(1)$, there is an $O(n^{2-4/p}/\alpha^4)$ dimension sketching algorithm, while for $p$ not an even integer, the $O(n^{2-4/p}/\alpha^4)$ dimension bound still holds if $\alpha$ is not too small. Our algorithm is based on a constant approximation algorithm for $\norm{A}_p$ when $p$ is an even integer.

\begin{lemma}[Theorem~8.2, \cite{lnw19}]
Suppose that $p$ is an even integer. There is a sketching algorithm whose output $Z$ satisfies $\norm{A}_p \le Z \le 2 \norm{A}_p$ with probability at least $2/3$. Furthermore, the sketching dimension of this algorithm is $O(n^{2-4/p})$.
\end{lemma}

\begin{algorithm}[t]
    \caption{$\alpha$-approximation for $\norm{A}_p$}
    \label{alg:schatten_p}
    \SetAlgoLined
    Set $q = p$ if $p \in 2\mathbb{Z}$, or to be the largest even integer less than $p$ otherwise\;
    $Let$ $\gA_q$ be a streaming algorithm that can output a constant-factor approximation to $\norm{A}_q$\;
    Set $t = (n^{1/2 - 1/p} / \alpha)^{1/(1/2 - 1/q)}$ \;
    $Let$ $G$ be an $r = t\poly\log(n/t) \times n$ matrix with i.i.d $\gN(0, 1/r)$ entries (these are i.i.d. normal random variables with mean $0$ and variance $1/r$) and let $H$ be an independent $r = O(t) \times n$ matrix with i.i.d $\gN(0, 1/r)$ entries.
    
    \ForEach{$\Delta_{i,j}$ in the data stream}{
        Compute the matrix $ G\Delta_{i, j}H^T$\;
        Add $G\Delta_{i,j}H^T$ to the input stream for $\gA_q$\;
    }
    Let $Z$ be the output from $\gA_q$\;
    \KwRet{Z}\;
\end{algorithm}

Our algorithm is given in Algorithm~\ref{alg:schatten_p}. For an even integer $p$, we maintain the matrix $GAH^T$ where $G$ and $H$ is defined in algorithm~\ref{alg:schatten_p} and use the constant approximation algorithm $\gA_q$ to estimate the Schatten-$q$ norm of $GAH^T$. The following lemma shows that $\norm{GA}_q$ can be an $\alpha$-approximation to $\norm{A}_p$.

\begin{lemma}[rewording of Theorem~22, \cite{li2017embeddings}]
Suppose that $p\geq q\geq 2$, $q$ is an even integer, and $t = O(n)$. Let $G$ be an $r \times n$ matrix with i.i.d.\ $\gN(0, 1/r)$ entries, where $r = O(t)$ when $q=2$ and $r=O(t\log^{1/(1/2-1/q)}(n/t))$ when $q\geq 4$.  Then, with probability at least $1 - \exp(c^\prime t)$, we have 
$\norm{A}_p \le \norm{\gamma GA}_q \le (n^{1/2 - 1/p})/(t^{1/2 - 1/q}) \norm{A}_p,$
where $\gamma$ is an appropriate scaling factor. 
\end{lemma}

If $H$ is a $(1/2)$-subspace embedding of $GA$, then we know that the singular values of $GAH^T$ are different from those of $GA$ by at most a constant factor (see Section~\ref{sec:prelim}), and thus $\norm{GAH^T}_q$ is a constant approximation to $\norm{GA}_q$. Recall that our sketch is a matrix of dimension $r\times O(t)$, where $r=t \poly(\log t)$, so the sketching dimension of our algorithm is $\tilde{O}(t^{2 - 4/p}) = \tilde{O}(n^{2 - 4/p}/\alpha^4)$.

\begin{theorem}
\label{thm:ub_schatten_even}
Suppose that $p\geq 2$ is an even integer. Then there is a sketching algorithm whose output $Z$ satisfies $\norm{A}_p \le Z \le \alpha \norm{A}_p$ with probability at least $2/3$. Furthermore, the sketching dimension of this algorithm is $\tilde{O}(n^{2-4/p}/\alpha^4)$.
\end{theorem}

When $p$ is not an even integer (and could even be a non-integer), let $q$ be the largest even integer that is smaller then $p$. Then our choice of $t$ still satisfies that $t = O(n)$ if $\alpha = \Omega(n^{1/p-1/q})$. Our arguments above continue to hold and we obtain the following theorem.

\begin{theorem}
\label{thm:ub_schatten_not_even}
Suppose that $p\geq 2$ is not an even integer. Let $q$ be the largest even integer less than $p$ and $\alpha = \Omega(n^{1/q-1/p})$. Then there is a sketching algorithm whose output $Z$ satisfies $\norm{A}_p \le Z \le \alpha \norm{A}_p$ with probability at least $2/3$. Furthermore, the sketching dimension of this algorithm is $\tilde{O}(n^{2-4/p}/\alpha^4)$.
\end{theorem}

\paragraph{Lower Bound.} Below we show that our upper bound is optimal up to $\mathrm{polylog}(n)$ factors. In~\cite{lnw19}, the authors give the following $n^{2}/\alpha^4$ lower bound for $\alpha$-approximating $\norm{A}_{\mathrm{op}}$.

\begin{lemma}[Corollary~3.3, \cite{lnw19}]
Suppose that $\alpha \ge 1 + c$ where $c$ is an arbitrarily small constant. Then, any sketching algorithm estimating $\norm{A}_{\mathrm{op}}$ within a factor $\alpha$ with failure probability smaller than $1/6$ requires sketching dimension $n^2 / \alpha^4$.
\end{lemma}

Since $\norminf{x} \le \norm{x}_p \le n^{1/p} \norminf{x}$, an $\alpha$-approximation of $\norm{A}_p$ implies an $\alpha n^{1/p}$ approximation to $\norminf{A} = \sigma_1(A)$. The following lower bound follows. 

\begin{theorem}
\label{thm:lb_schatten}
Suppose that $\alpha \ge 1 + c$, where $c > 0$ is an arbitrarily small constant. Then, any sketching algorithm estimating $\norm{A}_{\mathrm{p}}$ within a factor $\alpha$ with failure probability smaller than $1/6$ requires sketching dimension $O(n^{2-4/p} / \alpha^4)$.
\end{theorem}

\section{Cascaded Norms}
\label{sec:cascaded}
In this section, we consider approximating the cascaded $(p,q)$-norm of a matrix $X$, defined as $\norm{X}_{p,q} = (\sum_i (\sum_j |x_{ij}|^q)^{p/q})^{1/p}$, for a large approximation factor $\alpha$ when $p \ge 1$ and $q > 2$.
We follow the algorithm proposed in~\cite{andoni2011streaming}, which is based on the following precision sampling lemma.

\begin{lemma}[Precision Sampling~\cite{andoni2011streaming}]
\label{lem:nonUniformSampling}
Fix an integer $n\ge 2$, a multiplicative error parameter $\eps\in[1/n,1/3]$, and
an additive error parameter $\rho\in[1/n,1]$. Then there exist a
distribution $\mathcal{W}$ on the real interval $[1,\infty)$
and a 
reconstruction algorithm $R$, 
with the following two properties.
\begin{enumerate}[label = (\roman*)] 
\item (Accuracy)
Consider arbitrary $a_1,\ldots,a_n\in[0,1]$ and $f\in[1,1.5]$. 
Let $w_1,\ldots,w_n$ be chosen pairwise independently from $\mathcal{W}$. %
Then with probability at least $2/3$,
when algorithm $R$ is given $\{w_i\}_{i\in[n]}$ and $\{\hat a_i\}_{i\in[n]}$ 
such that each $\hat a_i$ is an arbitrary $(1/w_i,f)$-approximator of $a_i$,
it produces $\hat \sigma\ge 0$ which is
a $(\rho,f\cdot e^{\eps})$-approximator to $\sigma \eqdef \sum_{i=1}^n a_i$. Here, a {\em $(\rho,f)$-approximator} to $\tau>0$ is any quantity $\hat \tau$ 
satisfying
$
\tau/f-\rho\le \hat \tau\le f\tau+\rho.
$.
\item (Cost) There is $k=O(1/\rho\eps^2)$ such that 
the conditional expectation
$\E_{w\in \mathcal{W}}[{w \mid M}]\le O(k\log n)$ for some event $M=M(w)$
occurring with high probability. For every fixed $\alpha\in (0,1)$, we have $\E_{w\in
  \mathcal{W}}[{w^\alpha}]\le O(k^\alpha)$. 
The distribution
  $\mathcal{W}=\mathcal{W}(k)$ depends only on $k$.
\end{enumerate}
\end{lemma}

 When dealing with the cascaded norm $\norm{X}_{p,q}$, rather than storing in each hash bucket a number that is a weighted sum of different coordinates, the algorithm $\mathcal{A}$ in~\cite{andoni2011streaming} hashes rows into buckets, aggregates them, and stores in each bucket a sketch for the $\sigma_q$ estimation problem ($q$-th frequency moment) for the aggregated vector. In~\cite{andoni2011streaming}, the authors show the following theorem.

\begin{theorem}[rewording of Theorem 4.5 of \cite{andoni2011streaming}] 
For $p \ge 1, n \ge 2$, and $0 < \eps < 1/3$, let $L_q$ be a linear sketch: $\R^{d} \rightarrow \R^{S_q}$ with space $S_q = S_q(\alpha)$ such that there there is a reconstruction procedure $E_q: \R^{S_q} \rightarrow \R$ that can recover the $\ell_q$-norm of the input vector $x$ with an approximation factor $\alpha$.
Then there is a randomized algorithm that outputs an $\alpha$-approximation of $\norm{x}_{p, q}$ with probability at least $2/3$ using space $S \le S_q(\alpha/2)\cdot \beta(p, q)$. 
In particular, 
\begin{enumerate}[label = (\arabic*)]
    \item $\beta(p, q) \le O(n^{1-2/p} \cdot \left(pq\log n\right)^{O(1)})$ if $p, q > 2$.
    \item $\beta(p, q) \le O((pq\log n)^{O(1)})$ if $p 
    \le 2$ and $q > 2$.
\end{enumerate}
\end{theorem}

We note that the only difference with this version of the theorem and that in ~\cite{andoni2011streaming} is that the authors of ~\cite{andoni2011streaming} only consider sketches $L$ that can recover a $(1 \pm \eps)$-approximation to $\sigma_q$. In fact, the theorem above also holds for arbitrary $\alpha$-approximation sketches $L$. Notice that the algorithm we proposed in Section~\ref{sec:f_p} is actually a linear sketch. Hence, our theorem follows immediately.
\begin{theorem}
\label{thm:ub_cascaded}
Suppose that $\alpha\geq 8$. Then there is an algorithm whose output is $Z$, which satisfies that $\norm{X}_{p, q} \leq Z\leq \alpha\norm{X}_{p,q}$ with probability at least $2/3$. Furthermore, the algorithm uses $O(n^{1-2/p} d^{1-2/q} \cdot (pq\log n)^{O(1)} /\alpha^2)$ bits of space when $p, q > 2$ and uses $O(d^{1-2/q} \cdot (q\log n)^{O(1)} /\alpha^2)$ bits of space when $1 \le p < 2$ and $q > 2$.
\end{theorem}

\paragraph{Lower Bound.} We show that our upper bounds are tight up to $\poly(\log n)$ factors. First, we note that if $X$ contains nonzero entries in its first row only, then we have $\norm{X}_{p, q} = \norm{X_1}_q$, where $X_1$ is the first row of $X$. Hence, Theorem~\ref{thm:lb_fp} implies a space lower bound of $\Omega(d^{1 - 2/q}\log(M)\log(1/\delta) / \alpha^2)$ bits when $q > 2$.

For the case that $p, q > 2$, we consider the following modified version of the following $\ell_\infty^k $ communication problem (\cite{BJKS04}): there are two parties, Alice and Bob, holding matrices $X, Y \in \mathbb{Z}^{n \times d}$ respectively, and their goal is to decide if (i) $\max_{i,j}|X_{ij} - Y_{ij}| \le 1$ or (ii) $\max_{i,j}|X_{ij} - Y_{ij}| \ge k$. This problem requires $\Omega(nd/k^2)$ bits of total communication.

Let $k = 2 \alpha n^{1/p} d^{1/q}$. Note that for the case where $\max_{i,j} |X_{ij} - Y_{ij}|\le 1$, we have $\norm{X-Y}_{p,q}\le n^{1/p}d^{1/q}$. While for the case where $ \max_{i,j} |X_{ij} - Y_{ij}|\ge k$, we have $\norm{X-Y}_{p,q} \ge k = 2\alpha n^{1/p} d^{1/q}$. Suppose that an algorithm $\mathcal{A}$ can output a $Z$ such that $\norm{X}_{p,q} \le Z \le \alpha \norm{X}_{p,q}$ with probability at least $2/3$. Then we have the following protocol. Alice performs the update $X$ to the algorithm $\mathcal{A}$ and sends the memory of $\mathcal{A}$ to Bob, who then performs the update $-Y$ to $\mathcal{A}$ and outputs case (i) if $Z\leq \alpha n^{1/p}d^{1/q}$ and case (ii) otherwise. From the discussion above we know that Bob's output is correct with probability at least $2/3$, whence a lower bound of $\Omega(n^{1 - 2/p} d ^{1 - 2/q}/\alpha^2)$ bits follows. 

\begin{theorem}
\label{thm:lb_cascaded}
For the case that $p, q > 2$, any one-pass streaming algorithm that outputs a $Z$ such that $\norm{X}_{p,q} \le Z \le \alpha \norm{X}_{p,q}$ with probability at least $2/3$ requires $\Omega(n^{1 - 2/p} d^{1-2/q}/\alpha^2)$ bits of space. For the case that $1 \le p < 2$ and $q > 2$, any one-pass streaming algorithm that outputs a $Z$ such that $\norm{X}_{p,q} \le Z \le \alpha \norm{X}_{p,q}$ with probability at least $1 - \delta$ requires $\Omega(d^{1-2/q} \log(M) \log(1/\delta)/\alpha^{2})$ bits of space. 
\end{theorem}
\section{Rectangle $F_p$ $(p > 2)$}
\label{sec:rectangle}
In this section, we consider the rectangle $F_p$ problem. A rectangle-efficient algorithm was proposed in~\cite{tw12}. Instead of updating the counter in each coordinate inside a rectangle, they develop a rectangle-efficient data structure called \textsc{RectangleCountSketch}. We follow their notation that $O^\ast(f)$ denotes a function of the form $f\cdot \poly(\log(mn/\delta))$ for constant rectangle dimension $d$. 

\begin{lemma}[\textsc{RectangleCountSketch}, \cite{tw12}]
\label{lem:rectangle}
The data structure \textsc{RectangleCountSketch($\gamma$)}
can be updated rectangle-efficiently in time $O^\ast(\gamma^{-2})$. The total space is $O^\ast(\gamma^{-2})$ words. The data structure can be used to answer any query $i \in GF(n)^d$, returning a number $\Phi(i)$ with $|\Phi(i) - x_i| \le \gamma \|x\|_2$. The algorithm succeeds on all queries simultaneously with probability $\ge 1 - \delta$.
\end{lemma}

Based on the above data structure, in~\cite{tw12} the authors give an $O^\ast(n^{d(1-2/p)})$ space algorithm for the rectangle $F_p$ problem.
Similar to the algorithm in Section~\ref{sec:f_p}, let $n^{d(1-2/q)} = n^{d(1-2/p)}/\alpha^2$. Then $\norm{x}_q$ can be an $\alpha$-approximation of $\norm{x}_p$. Under a similar analysis we obtain the following theorem.

\begin{theorem}\label{thm:rectangle}
Suppose that $p>2$. There is a rectangle-efficient one-pass streaming algorithm which outputs a number $Z$ that is an $\alpha$-approximation to $\norm{x}_p^p$, i.e., $\norm{x}_p^p \le Z \le \alpha \norm{x}_p^p$, with probability at least $1 - \delta$. It uses $O^\ast(n^{d(1-2/p)} / \alpha^{2/p})$ bits of space and $O^\ast(n^{d(1-2/p)} / \alpha^{2/p})$ time to process each rectangle in the stream.
\end{theorem}

\section*{Acknowledgements.} Yi Li would like to thank for partial support from the Ministry of Education of Singapore under a Tier 1 Grant RG75/21.  Honghao Lin and David Woodruff would like to thank for partial support from the National Science Foundation (NSF) under Grant No. CCF-1815840.

\bibliography{reference}
\bibliographystyle{alpha}  

\appendix
\section{Proof of Lemma~\ref{lem:new_bounds_bin}}\label{sec:concentration Y_{-I}}

The first result, Equation~\eqref{new_eqn:E Y_{-I}}, was proved in~\cite{WW15}. Now we prove the second result.

By a standard symmetrization technique (see, e.g., \cite[p153]{LT91}),
\begin{equation}\label{eqn:Y_{-I} aux1}
\left(\E \abs{ \norm{\bY_{-I}}_p^p - \E\left[\norm{\bY_{-I}}_p^p\right] }^r\right)^{\frac{1}{r}} = \left(\E \abs{\sum_{i\neq I} (\bY_i^p -\E\bY_i^p)}^r\right)^{\frac{1}{r}} 
\leq 2 \left(\E \abs{\sum_{i\neq I} \eps_i \bY_i^p}^r \right)^{\frac{1}{r}},
\end{equation} \\ multline
where the $\eps_i$ are independent Rademacher variables.

By Lata\l{}a's inequality (\cite[Corollary 3]{L97}), it holds for $r\geq 2$ that
\begin{equation}\label{eqn:latala}
\left(\E \abs{\sum_{i\neq I} \eps_i \bY_i^p}^r \right)^{\frac{1}{r}} \leq K_1\frac{r}{\ln r}\max\left\{ \left(\E \sum_{i\neq I} \bY_i^{2 p}\right)^{\frac12}, \left(\E \sum_{i\neq I} \bY_i^{r p}\right)^{\frac{1}{r}}  \right\},
\end{equation}
where $K_1 > 0$ is an absolute constant.

It was shown in~\cite[Lemma 6.3]{WW15} that
\[
\E \bY_i^p \leq K_2^p p^p 10^{T p},\quad p\geq 1,
\]
for some absolute constant $K_2 > 0$. It then follows that
\begin{equation}\label{eqn:Y_{-I} aux2} 
\left(\E \sum_{i\neq I} \bY_i^{rp}\right)^{\frac{1}{r}} \leq K_2^p (r p)^p n^{1/r} 10^{p T}
\end{equation}
The result follows from combining~\eqref{eqn:Y_{-I} aux1}, \eqref{eqn:latala} and \eqref{eqn:Y_{-I} aux2}.

\bigskip

\noindent\textit{Remark.} We note an omission in~\cite{WW15}. In that paper, the proof of the second result, i.e.,\  Equation~\eqref{eqn:Y_{-I} special moment}, assumes that the larger term in~\eqref{eqn:latala} is $(\E\sum_{i\neq I} \bY_i^{2r})^{1/p}$, which is not necessarily the case. Lemma 2.5 in that paper is also an incorrect citation from~\cite{L97}, since the conclusion should be $\max\{\Delta_1(X), \Delta_\ell(X)\}$ for nonnegative variables $X$, but this would be too large for the proof. Hence we first symmetrize the variables, which allows for a better bound on $\max\{\Delta_2(X), \Delta_\ell(X)\}$.
\section{Omitted Proofs in Section~\ref{sec:L0}}\label{sec:l0_proofs}

To prove Lemma~\ref{lem:bucket_value}, we need an auxiliary lemma, which is a corollary of the Paley-Zygmund inequality.

\begin{lemma}[Second moment method]\label{lem:second moment method}
Let $X \geq 0$ be a random variable (not identically $0$) with finite variance, then
$
\Pr[X > 0] \geq (\E X)^2/(\E X^2)
$.
\end{lemma}

Now we are ready to show Lemma~\ref{lem:bucket_value}.
\begin{proof}[Proof of Lemma~\ref{lem:bucket_value}]
Observe that $\E X^2 = \sum_i v_i^2 > 0$ and $\E X^4 = \sum_i v_i^4 + 3 \sum_{i \neq j} v_i^2 v_j^2 \leq 3(\E X^2)^2$. It follows from Lemma~\ref{lem:second moment method} that
$\Pr[X = 0] = \Pr[X^2 = 0] \leq 1 - (\E X^2)^2/(\E X^4) \leq 2/3$. 
\end{proof}

Lemma~\ref{lem:length of x} follows from a straightforward calculation.
\begin{proof}[Proof of Lemma~\ref{lem:length of x}]
Note that $l = t/8 \leq (\log n)/8$. The length of $x$ equals $\sum_{i=1}^{l} \frac{1}{2}\left\lceil n^{i/l} \right\rceil \leq \frac{l}{2} + \frac{1}{2}n^{1/l} \frac{n - 1}{n^{1/l} - 1} \leq n$ when $n$ is large enough.
\end{proof}
\section{Two-Pass $\ell_0$ Estimation Algorithm}\label{sec:L0_multipass}

Recall the one-pass algorithm in \cite{knw10b}, which downsamples the $n$ coordinates in $\log n$ levels $S^1,\dots,S^{\log n}$.
There exists a level $J$ satisfying $1\leq \E[\ell_0(S^{J})]\leq 2$. Then $2^j$ is a constant-factor estimation for $j=J+O(1)$. For $j=J+O(1)$, it can be shown that $\ell_0(S^j)$ concentrates around the expectation and will not exceed a constant $c$ with high probability. Hence, the coordinates in level $j$ are perfectly hashed with high probability and the exact $\ell_0(S^j)$ can be obtained. Overall, the algorithm outputs $J+O(1)$ up to an additive $O(1)$ error and obtains a constant-factor approximation with constant probability. 

Denote by $\tilde{\ell}_0$ the constant factor approximation. In light of Lemma~\ref{lemma:balls to bins with limited independence}, it examines the $( \log (16\tilde{\ell}_0 \eps^{2}))$-th level and counts the number $T$ of bins that receive at least one ball in that level. It then shows that $32 \tilde{\ell}_0\eps^2 \ln(1-T\eps^2)/\ln(1-\eps^2)$ is a $(1 \pm O(\eps))$-estimation. By rescaling $O(\eps)$ to $\eps$ we obtain a $(1 \pm \eps)$-estimation.

\begin{lemma}[balls to bins with limited independence~{\cite{knw10b}}]\label{lemma:balls to bins with limited independence}
    Throw $A$ balls into $K$ bins using a hash function $h \in \mathcal{H}_A([A], [K])$. Let $X$ be the number of bins receiving at least one ball. Alternatively, use a hash function $h' \in \mathcal{H}_k([A], [K])$, where $k = \Omega(\log (1/\epsilon)/\log\log (1/\epsilon))$. Let $X'$ be the number of bins receiving at least one ball in this case. Then there exists a constant $\epsilon_0$ such that the following holds: if $100 \leq A \leq K/20$ with $K = 1/\epsilon^2$ and $\epsilon \leq \epsilon_0$, then we have
    \[
        \Pr[|X' - \E[X]| \leq 8\epsilon \E[X]] \geq 4/5.
    \]
\end{lemma}

The key observation for us is the following: if we know a constant-factor approximation, we only need to maintain one level. Moreover, if we use Theorem~\ref{thm:n^1/t-approximator} to obtain a $(\log M)$-factor approximation, we will only need to maintain $O(\log \log M)$ levels. 

\begin{algorithm}[t]
    \caption{$(1 \pm \eps)$-approximator for $\ell_0$}\label{alg:2-pass-new}
    \SetAlgoLined

    Set $K = 1/\eps^{2}$\;
    Set $B = \log \tilde{\ell}_0' - \log \eps^{-2}$\;
    Set $R = \log\log M$\;
    Initialize $K \cdot (2R + 21)$ counters $C_{-R-10,1},\dots,C_{R+10,K}$ to $0$\;
    Pick random hash function $h_1 \in \mathcal{H}_2([n], [0, n-1])$, $h_2 \in \mathcal{H}_2([n], [K^3])$, $h_3 \in \mathcal{H}_k([K^3], [K])$ where $k = \Omega(\log(1/\eps)/\log\log (1/\eps))$, $h_4 \in \mathcal{H}_2([K^3], [K])$\;
    Pick a prime $p \in [D, D^3]$ where $D = 100K\log M$\;
    Pick a random vector $\boldsymbol{u} \in \mathbb{F}^K_p$\;

    \ForEach{$(x, v)$ in the data stream}{
        $b \gets \operatorname{lsb}(h_1(x)) - B$\;
        \For{$i \gets -B - R -10 $ \KwTo $\max(b, B + R + 10)$}{
            $C_{i, h_3(h_2(x))} \gets (C_{i, h_3(h_2(x))} + v \cdot \boldsymbol{u}_{h_4(h_2(x))}) \bmod p$\;
        }
    }
    
    $J \gets$ largest $j$ such that $T_j = |\{k \mid C_{j, k} > 0\}| > 0.011K$\;

    \KwRet{$2^{J + B} \ln(1-T_J/K)/\ln(1-1/K)$}\;
\end{algorithm}

Our algorithm is presented in Algorithm~\ref{alg:2-pass-new}, assuming in the first-pass that we obtain a $(\log M)$-approximation $\tilde{\ell}_0'$ as in the algorithm above, using $O(\log n + \frac{\log M}{\log \log M} \cdot \log\log M) = O(\log n)$ bits. In the second pass, we maintain $O(\log \log M)$ levels of the sampling scheme and examine the deepest level of $\Theta(\eps^{-2})$ survivors. To show the correctness of our algorithm, we need the following lemmas.

\begin{lemma}[\cite{knw10b}]\label{lemma:good J}
There exists an absolute constant $\eps_0 > 0$ such that the following holds.
Suppose that $\eps\in(0,\eps_0)$ and $1/(300\eps^2) \leq \ell_0(S^J) \leq 1/(20\eps^2)$ for some level $J$. Let $T$ be the number of bins receiving at least one ball under the process in Lemma \ref{lemma:balls to bins with limited independence} with $K=\eps^{-2}$. With probability at least $0.95-O(\eps^2)$, the quantity $2^J \ln(1-T\eps^2)/\ln(1-\eps^2)$ is a $(1 \pm O(\eps))$-approximation to $\ell_0$. The entire process can be maintained using $O(\log n + \eps^{-2}(\log(1/\eps)+ \log\log M))$ bits of space.
\end{lemma}

\begin{proof}
The lemma is implicit in Lemma \ref{lemma:balls to bins with limited independence} and Theorem 3 in \cite{knw10b}. By adjusting constants, we can amplify the success probability to $0.95-O(\eps^2)$.
\end{proof}

\begin{lemma}\label{lem:two-pass big ell_0}
Assume that $\ell_0 > 1/(32\eps^2)$. Algorithm~\ref{alg:2-pass-new} find a $J$ such that $1/(300\eps^2) \leq \ell_0(S^J) \leq 1/(20\eps^2)$ with probability at least $0.85-O(\eps^2)-\exp(-\Omega(1/\eps^2))$ using $O(\log n + \eps^{-2}\log\log M(\log(1/\eps) + \log\log M))$ bits of space.
\end{lemma}

\begin{proof}
The main technique is from Lemma 4 in \cite{knw10b}. We subsample the data stream into $\log n$ levels, and in each level hash the surviving coordinates into $K = 1/\eps^2$ buckets. The algorithm outputs the largest level $j$ such that the number of bins receiving at least one ball is at least $0.011/\eps^2$.

Note that $\mathbf{Var}[\ell_0(S^{j})] \leq \E[\ell_0(S^{j})]$. By Chebyshev's inequality, it holds for each $j$ that
\[
\Pr[|\ell_0(S^j) - \E \ell_0(S^j)| > c\E \ell_0(S^j)] \leq 1/(c^2\E \ell_0(S^j)).
\]

There exists a level $j^\ast$ such that $1/(64\eps^2) \leq \E[\ell_0(S^{j^{\ast}})] \leq 1/(32\eps^2)$, so by Chebyshev's inequality, $j^\ast$ and $j^\ast + 1$ are both good $J$.  Note that levels in $\log \tilde{\ell}_0' - \log \eps^{-2} \pm \log\log M$ must contain a level with expectation in $[1/\eps^2, 2/\eps^2]$, and the constant factor before $1/\eps^2$ in the expectation of a good $J$ is at most $128$. Thus,  maintaining ten additional levels is enough to contain the levels we want. The algorithm maintains the levels in $(\log \tilde{\ell}_0' - \log \eps^{-2} \pm (\log\log M + 10))$, to ensure containing $j^\ast$ and $j^\ast + 1$ given $\tilde{\ell}_0'$.

If there is a significant difference between the result of the process of balls to bins for $j^\ast$ and $j > j^\ast + 1$ with appreciable probability, then we can detect this, and the algorithm will output correctly. This is what we prove in the following.

By Chebyshev's inequality, taking $c_1 = 1/5$ and $c_2 = 3/5$ we have
\begin{multline*}
\Pr\left[\ell_0(S^{j^{\ast}}) < 1/(80\eps^2) \right] 
\leq \Pr\left[|\ell_0(S^{j^\ast}) - \E \ell_0(S^{j^\ast})| \geq c_1\E \ell_0(S^{j^\ast})\right] 
\leq (64/c_1^2) \eps^2 = 1600\eps^2 
\end{multline*}
and
\begin{multline*}
\Pr\left[\ell_0(S^{j^{\ast}}) > 1/(20\eps^2) \right] 
\leq \Pr\left[|\ell_0(S^{j^\ast}) - \E \ell_0(S^{j^\ast})| \geq c_2\E \ell_0(S^{j^\ast})\right] 
\leq (64/c_2^2) \eps^2 \leq 178\eps^2.
\end{multline*}
Combining the two results, we have
\begin{equation}\label{eqn:event ell_0(S^j*)}
\Pr\left[ 1/(80\eps^2) \leq \ell_0(S^{j^{\ast}}) \leq 1/(20\eps^2)\right] \geq  1 - 1778\eps^2.
\end{equation}
Furthermore, for each $j > j^\ast + 1$, note that $1/(256\cdot 2^{j-j^\ast-2} \eps^2) \leq \E[\ell_0(S^{j})] \leq 1/(128 \cdot 2^{j-j^\ast-2}\eps^2)$. Thus, we have
\begin{multline*}
    \Pr\left[\ell_0(S^{j}) > 1/(100\eps^2)\right]
    \leq \Pr\left[|\ell_0(S^{j}) - \E \ell_0(S^{j})| > 1/(5\cdot 2^{j-j^\ast-2})\cdot \E \ell_0(S^{j})\right] 
    \leq (25 \cdot 256 \eps^2)/2^{j - j^\ast - 2}
\end{multline*}
By a union bound, we see that $\ell_0(S^{j}) \leq 1/(100\eps^2)$ for all $j > j^\ast + 1$ with probability at least
\begin{equation}
\label{eqn:event L_j}
\Pr\left[ \ell_0(S^{j}) \leq 1/(100\eps^2)\text{ for all }j > j^\ast + 1 \right] 
\geq 1 -
\sum_{j > j^\ast + 1} \frac{25 \cdot 256 \eps^2}{2^{j - j^\ast - 2}}
\geq 1 - 25 \cdot 512 \eps^2 = 1 - 12800 \eps^2.
\end{equation} 

Next we bound the probability that each bucket is correctly maintained. Let $\cE$ be the event that $S^{j^\ast}$ is perfectly hashed by $h_2$. Then $\Pr[\cE] = 1 - O(1/K)$ since the range of $h_2$ is of size $K^3$, and $\cE$ still holds for $j > j^\ast$, since we use the same function and $S^j \subseteq S^{j^\ast}$ for $j > j^\ast$.

Conditioned on $\cE$ and $\ell_0(S^{j^\ast}) \leq K/20$ occurring, we define two events as per Lemma 6 of \cite{knw10b}. Define the event $\cQ$ to be that $p$ does not divide the frequency of any coordinate in $S^{j^\ast}$, and the event $\cQ'$ to be the event that $h_4(h_2(i)) \neq h_4(h_2(i))$ for distinct $i, i' \in S^{j^\ast}$ with $h_3(h_2(i)) = h_3(h_2(i'))$. Since $S^j \subseteq S^{j^\ast}$ for $j > j^\ast$, these events still hold for $j > j^\ast$.

We have that $\Pr[\cQ] = 1-O(1/K)$ and $\Pr[\cQ'] > 7/8$. By Fact 3 in \cite{knw10b}, conditioned on $\cQ$ and $\cQ'$, each bucket is correctly maintained with probability $1/p$. Taking a union bound over all buckets in levels $j \in \{j^\ast, j^\ast + 1\}$ and noticing that $2K / p \leq 1/50$, we see that all buckets in levels $j \in \{j^\ast, j^\ast + 1\}$ are correctly maintained with probability at least $7/8 \cdot 49/50 - O(1/K)$.

Let $T_j$ be the number of bins receiving at least one ball in the $j$-th level. Then 
\[
\E[T_j] = K \left(1-(1-1/K)^{\ell_0(S^j)} \right).
\]
Next we condition on the events in \eqref{eqn:event ell_0(S^j*)} and \eqref{eqn:event L_j}, and condition on buckets in level $j \in \{j^\ast, j^\ast + 1\}$ being correctly maintained. We do not need to consider levels $j > j^\ast + 1$, because the error only makes $T_j$ smaller, which does not effect our criterion.

Since $(1-1/x)/e \leq (1-1/x)^x \leq 1/x$ for all $x \geq 1$, we have (setting $x=1/\eps^2$)
\[
\E[T_j]/K \in \left[ 1-e^{-\ell_0(S^j)/K}, (1 - e^{-\ell_0(S^j)/K}(1-1/K)^{\ell_0(S^j)/K}) \right]
\]
Recall that $K = 1/\eps^2$ and $\eps \leq 1/10$. We thus have 
\begin{align*}
\E[T_{j^\ast}] &\geq (1 - e^{-1/80})/\eps^2 > 0.012/\eps^2, \\
\E[T_j] &\leq (1 - e^{-1/100}(1-\eps^2)^{1/100})/\eps^2 < 0.011/\eps^2, \quad j > j^\ast + 1.
\end{align*}
Therefore, a gap exists between $\E[T_{j^\ast}]$ and $\E[T_{j}]$ for $j > j^\ast + 1$. 

By negative dependence in the process of balls to bins, we have the following Chernoff bounds on $T_j$~\cite{dubhashi1996balls}:
\begin{align*}
\Pr[T_{j^\ast} > 0.011/\eps^2] &= 1-e^{-\Omega(1/\eps^2)} \\
\Pr[T_{j} < 0.011/\eps^2] &= 1-e^{-\Omega(1/\eps^2)}, \quad \forall j > j^\ast + 1.
\end{align*}
Overall, the algorithm will output $J\in \{j^\ast, j^\ast + 1\}$ which satisfies that $1/(300\eps^2) \leq \ell_0(S^J) \leq 1/(20\eps^2)$ with probability at least $0.85 - O(\eps^2)-\exp(-\Omega(1/\eps^2))\geq 0.85 - O(\eps^2)$ for $\eps$ small enough.
\end{proof}

If $\ell_0$ is small, $J$ in Lemma \ref{lemma:good J} may not exist. When $\ell_0\leq 1/(32\eps^2)$, we can invoke the algorithm in Theorem 4 and Lemma 6 of~\cite{knw10b} instead. Adjusting the constants, we have the following lemma.

\begin{lemma}\label{lem:two-pass small ell_0}
There exists an algorithm that outputs a $1 \pm O(\eps)$-approximation to $\ell_0$ if $\ell_0 \leq 1/(32\eps^2)$ and otherwise outputs $-1$ with probability at least $0.85 - O(\eps^2)$. It uses  $O(\log n + \eps^{-2}(\log(1/\eps) + \log\log M))$ bits of space.
\end{lemma}

We run the algorithms in Lemmas~\ref{lem:two-pass small ell_0} and~\ref{lem:two-pass big ell_0} in parallel. If the former algorithm outputs $-1$, we take the output of the latter one as the final output. The next theorem follows from Lemmas~\ref{lemma:good J}, \ref{lem:two-pass big ell_0} and~\ref{lem:two-pass small ell_0}, where the success probability is at least $0.65 - O(\eps^2)\geq 0.6$, provided that $\eps$ is small enough. Theorem~\ref{thm:two-pass} then follows.
    

\end{document}